\documentclass[12pt, a4paper]{article}
\usepackage{graphicx,epsfig}

\begin{document}

\title{On the Cosmological Constant Problems and the Astronomical
Evidence for a Homogeneous Energy Density with Negative Pressure 
\footnote{Invited lecture at the first {\it S\'{e}minaire Poincar\'{e}}, 
Paris, March 2002.} }

\author{Norbert Straumann\\
        Institute for Theoretical Physics University of Zurich,\\
        CH--8057 Zurich, Switzerland}
\date{\today}

\maketitle

\begin{abstract}
In this article the cosmological constant problems, as well as the
astronomical evidence for a cosmologically significant homogeneous
exotic energy density with negative pressure (quintessence), are
reviewed for a broad audience of physicists and mathematicians.
After a short history of the cosmological term it is explained why
the (effective) cosmological constant is expected to obtain
contributions from short-distance physics corresponding to an
energy scale at least as large as the Fermi scale. The actual tiny
value of the cosmological constant by particle physics standards
represents, therefore, one of the deepest mysteries of present-day
fundamental physics. In a second part I shall discuss recent
astronomical evidence for a cosmologically significant vacuum
energy density or an effective equivalent, called quintessence.
Cosmological models, which attempt to avoid the disturbing cosmic
coincidence problem, are also briefly reviewed.
\end{abstract}

\section{Introduction}

In recent years important observational advances have led quite
convincingly to the astonishing conclusion that the present
Universe is dominated by an exotic homogeneous energy density with
{\it negative} pressure. I shall discuss the current evidence for
this unexpected finding in detail later on, but let me already indicate in this
introduction the most relevant astronomical data.

First, we now have quite accurate measurements of the anisotropies
of the cosmic microwave background radiation (CMB). In particular,
the position of the first acoustic peak in the angular power
spectrum implies that the Universe is, on large scales, nearly
flat (Sect.6).

On the other hand, a number of observational results, for instance
from clusters of galaxies, show consistently that the amount of
``matter'' (baryons and dark matter) which clumps in various
structures is significantly {\it undercritical}. Hence, there must
exist a {\it homogeneously} distributed exotic energy component.

Important additional constraints come from the Hubble diagram of
type Ia supernovas at high redshifts. Although not yet as
convincing, they support these conclusions (Sect.5). More recently,
the combination of CMB data and information provided by large scale
galaxy redshift surveys have given additional confirmation.

Some of you may say that all this just shows that we have to keep
the cosmological term in Einstein's field equations, a possibility
has been considered during all the history of relativistic
cosmology (see Sect.2). From our present understanding we would
indeed expect a non-vanishing cosmological constant, mainly on the
basis of quantum theory, as will be discussed at length later on.
However, if a cosmological term describes the astronomical
observations, then we are confronted with two difficult problems,
many of us worry about:

The first is the {\it old mystery}: Since all sorts of vacuum
energies contribute to the effective cosmological constant (see
Sect.4), we wonder why the total vacuum energy density is so
incredibly small by all particle physics standards. Theoreticians
are aware of this profound problem since a long time, -- at least
those who think about the role of gravity among the fundamental
interactions. Most probably, we will only have a satisfactory
answer once we shall have a theory which successfully combines the
concepts and laws of general relativity about gravity and
spacetime structure with those of quantum theory.

Before the new astronomical findings one could at least hope that
we may one day have a basic understanding for a vanishing
cosmological constant, and there have been interesting attempts in
this direction (see, e.g., Ref. \cite{1}). But now we are also
facing a {\it cosmic coincidence} problem: Since the vacuum energy
density is constant in time -- at least after the QCD phase
transition --, while the matter energy density decreases as the
Universe expands, it is more than surprising that the two are
comparable just at the present time, while their ratio has been
tiny in the early Universe and will become very large in the
distant future.

This led to the idea that the effective cosmological constant we
observe today is actually a {\it dynamical} quantity, varying with
time. I want to emphasize already now that these so-called {\it
quintessence} models do, however, not solve the first problem.
(More on this in Sect.7.)

This paper is organized as follows. Section 2 is devoted to the instructive 
early history of the $\Lambda-$term, including some early remarks by Pauli 
on the quantum aspect connected with it. In Section 3 we recall important examples
of vacuum energies in quantum electrodynamics and their physical significance 
under variable external conditions. We then shown in Section 4 that simple and 
less naive order of magnitude estimates of various contributions to 
the vacuum energy density of the Standard Model all lead to expectations
which are in gigantic conflict with the facts. I then turn to the astronomical 
and astrophysical aspects of our theme. In Section 5 it will be described what is known
about the luminosity-redshift relation for type Ia supernovas. The remaining
systematic uncertainties are discussed in some detail. Most space of Section 6
is devoted to the physics of the CMB, including of how the system of basic equations
which govern its evolution before and after recombination is obtained. We then 
summarize the current observational results, and what has been learned from them
about the cosmological parameters. We conclude in Section 7 with a few remarks about
the goal of quintessence models and the main problems this scenario 
is facing.

\section{On the history of the $\Lambda$-term}

The cosmological term was introduced by Einstein when he applied
general relativity for the first time to cosmology. In his paper
of 1917 \cite{2} he found the first cosmological solution of a
consistent theory of gravity. In spite of its drawbacks this bold
step can be regarded as the beginning of modern cosmology. It is
still interesting to read this paper about which Einstein says:
{\it ``I shall conduct the reader over the road that I have myself
travelled, rather a rough and winding road, because otherwise I
cannot hope that he will take much interest in the result at the
end of the journey.''} In a letter to P. Ehrenfest on 4 February
1917 Einstein wrote about his attempt: {\it ``I have again
perpetrated something relating to the theory of gravitation that
might endanger me of being committed to a madhouse. (Ich habe
wieder etwas verbrochen in der Gravitationstheorie, was mich ein
wenig in Gefahr bringt, in ein Tollhaus interniert zu werden.)''}
\cite{3}.

In his attempt Einstein assumed -- and this was completely novel
-- that space is globally {\it closed}, because he then believed
that this was the only way to satisfy Mach's principle, in the
sense that the metric field should be determined uniquely by the
energy-momentum tensor. In addition, Einstein assumed that the
Universe was {\it static}. This was not unreasonable at the time,
because the relative velocities of the stars as observed were
small. (Recall that astronomers only learned later that spiral
nebulae are independent star systems outside the Milky Way. This
was definitely established when in 1924 Hubble found that there
were Cepheid variables in Andromeda and also in other galaxies.
Five years later he announced the recession of galaxies.)

These two assumptions were, however, not compatible with
Einstein's original field equations. For this reason, Einstein
added the famous $\Lambda$-term, which is compatible with the
principles of general relativity, in particular with the
energy-momentum law $\nabla_\nu T^{\mu\nu}=0$ for matter. The
modified field equations in standard notation (see, e.g., \cite{15}) 
and signature $(+---)$ are
\begin{equation}
G_{\mu\nu} = 8\pi G T_{\mu\nu} + \Lambda g_{\mu\nu}.
\end{equation}
For the static Einstein universe these equations imply the two
relations
\begin{equation}
8\pi G \rho = \frac{1}{a^2} = \Lambda,
\end{equation}
where $\rho$ is the mass density of the dust filled universe (zero
pressure) and $a$ is the radius of curvature. (We remark, in
passing, that the Einstein universe is the only static dust
solution; one does not have to assume isotropy or homogeneity. Its
instability was demonstrated by Lema\^{\i}tre in 1927.) Einstein
was very pleased by this direct connection between the mass
density and geometry, because he thought that this was in accord
with Mach's philosophy. (His enthusiasm for what he called Mach's
principle later decreased. In a letter to F.Pirani he wrote in
1954: {\it ``As a matter of fact, one should no longer speak of
Mach's principle at all. (Von dem Machschen Prinzip sollte man
eigentlich \"uberhaupt nicht mehr sprechen''.)} \cite{4})

In the same year, 1917, de Sitter discovered a completely
different static cosmological model which also incorporated the
cosmological constant, but was {\it anti-Machian}, because it
contained no matter \cite{5}. The model had one very interesting
property: For light sources moving along static world lines there
is a gravitational redshift, which became known as the {\it de
Sitter effect}. This was thought to have some bearing on the
redshift results obtained by Slipher. Because the fundamental
(static) worldlines in this model are not geodesic, a freely-
falling particle released by any static observer will be seen by
him to accelerate away, generating also local velocity (Doppler)
redshifts corresponding to {\it peculiar velocities}. In the
second edition of his book \cite{6}, published in 1924, Eddington
writes about this:
\begin{quote}
{\it ``de Sitter's theory gives a double explanation for this motion
of recession; first there is a general tendency to scatter (...);
second there is a general displacement of spectral lines to the
red in distant objects owing to the slowing down of atomic
vibrations (...), which would erroneously be interpreted as a
motion of recession.''}
\end{quote}
I do not want to enter into all the confusion over the de Sitter
universe. This has been described in detail elsewhere (see, e.g.,
\cite{7}). An important discussion of the redshift of galaxies in
de Sitter's model by H. Weyl \cite{8} in 1923 should, however, be
mentioned. Weyl introduced an expanding version of the de Sitter
model\footnote{I recall that the de Sitter model has many
different interpretations, depending on the class of fundamental
observers that is singled out.}. For {\it small} distances his
result reduced to what later became known as the Hubble law.

Until about 1930 almost everybody {\it knew} that the Universe was
static, in spite of the two fundamental papers by Friedmann
\cite{9} in 1922 and 1924 and Lema\^{\i}tre's independent work
\cite{10} in 1927. These path breaking papers were in fact largely
ignored. The history of this early period has -- as is often the
case -- been distorted by some widely read documents. Einstein too
accepted the idea of an expanding Universe only much later. After
the first paper of Friedmann, he published a brief note claiming
an error in Friedmann's work; when it was pointed out to him that
it was his error, Einstein published a retraction of his comment,
with a sentence that luckily was deleted before publication: {\it
``[Friedmann's paper] while mathematically correct is of no physical
significance''}. In comments to Lema\^{\i}tre during the Solvay
meeting in 1927, Einstein again rejected the expanding universe
solutions as physically unacceptable. According to Lema\^{\i}tre,
Einstein was telling him: {\it ``Vos calculs sont corrects, mais
votre physique est abominable''}. On the other hand, I found in the
archive of the ETH many years ago a postcard of Einstein to Weyl
from 1923 with the following interesting sentence: {\it ``If there
is no quasi-static world, then away with the cosmological term''}.
This shows once more that history is not as simple as it is often
presented.

It also is not well-known that Hubble interpreted his famous
results on the redshift of the radiation emitted by distant
`nebulae' in the framework of the de Sitter model. He wrote:
\begin{quote}
{\it ``The outstanding feature however is that the velocity-distance
relation may represent the de Sitter effect and hence that
numerical data may be introduced into the discussion of the
general curvature of space. In the de Sitter cosmology,
displacements of the spectra arise from two sources, an apparent
slowing down of atomic vibrations and a tendency to scatter. The
latter involves a separation and hence introduces the element of
time. The relative importance of the two effects should determine
the form of the relation between distances and observed
velocities.''}
\end{quote}
 However, Lema\^{\i}tre's successful explanation of Hubble's
discovery finally changed the viewpoint of the majority of workers
in the field. At this point Einstein rejected the cosmological
term as superfluous and no longer justified  \cite{11}. He
published his new view in the {\it Sitzungsberichte der
Preussischen Akademie der Wissenschaften.} The correct citation
is:

 \begin{center}
 Einstein. A. (1931). Sitzungsber. Preuss. Akad. Wiss. 235-37.
 \end{center}

Many authors have quoted this paper but never read it. As a
result, the quotations gradually changed in an interesting, quite
systematic fashion. Some steps are shown in the following
sequence:
 \begin{enumerate}
 \item[-]{A. Einstein. 1931. Sitzsber. Preuss. Akad. Wiss. ...}
 \item[-]{A. Einstein. Sitzber. Preuss. Akad. Wiss. ... (1931)}
 \item[-]{A. Einstein (1931). Sber. preuss. Akad. Wiss. ...}
 \item[-]{Einstein. A .. 1931. Sb. Preuss. Akad. Wiss. ...}
 \item[-]{A. Einstein. S.-B. Preuss. Akad. Wis. ...1931}
 \item[-]{A. Einstein. S.B. Preuss. Akad. Wiss. (1931) ...}
 \item[-]{Einstein, A., and Preuss, S.B. (1931). Akad. Wiss. \textbf{235}}
 \end{enumerate}

Presumably, one day some historian of science will try to find out
what happened with the young physicist S.B. Preuss, who apparently
wrote just one important paper and then disappeared from the
scene.

Einstein repeated his new standpoint much later  \cite{12}, and
this was also adopted by many other influential workers, e.g., by
Pauli  \cite{13}. Whether Einstein really considered the introduction of
the $\Lambda$-term as ``the biggest blunder of his life'' appears
doubtful to me. In his published work and letters I never found
such a strong statement. Einstein discarded the cosmological term
just for simplicity reasons. For a minority of cosmologists
(O.Heckmann, for example  \cite{14}), this was not sufficient
reason.

After the $\Lambda$-force was rejected by its inventor, other
cosmologists, like Eddington, retained it. One major reason was
that it solved the problem of the age of the Universe when the
Hubble time scale was thought to be only 2 billion years
(corresponding to the value $H_0 \sim 500\ km\ s^{-1} Mpc^{-1}$ of
the Hubble constant). This was even shorter than the age of the
Earth. In addition, Eddington and others overestimated the age of
stars and stellar systems.

For this reason, the $\Lambda$-term was employed again and a model
was revived which Lema\^{\i}tre had singled out from the many
solutions of the Friedmann-Lema\^{\i}tre equations\footnote{I
recall that Friedmann included the $\Lambda$-term in his basic
equations. I find it remarkable that for the negatively curved
solutions he pointed out that these may be open or compact (but
not simply connected).}. This so-called Lema\^{\i}tre hesitation
universe is closed and has a repulsive $\Lambda$-force
($\Lambda>0$), which is slightly greater than the value chosen by
Einstein. It begins with a big bang and has the following two
stages of expansion. In the first the $\Lambda$-force is not
important, the expansion is decelerated due to gravity and slowly
approaches the radius of the Einstein universe. At about the same
time, the repulsion becomes stronger than gravity and a second
stage of expansion begins which eventually inflates into a
whimper. In this way a positive $\Lambda$ was employed to
reconcile the expansion of the Universe with the age of stars.

The repulsive effect of a positive cosmological constant can be
seen from the following consequence of Einstein's field equations
for the time-dependent scale factor $a(t)$:
\begin{equation}
\ddot{a} = -\frac{4\pi G}{3}(\rho + 3p)a + \frac{\Lambda}{3}a,
\end{equation}
where $p$ is the pressure of all forms of matter.

Historically, the Newtonian analog of the cosmological term was
regarded by Einstein, Weyl, Pauli, and others as a {\it{Yukawa
term}}. This is not correct, as I now show.

For a better understanding of the action of the $\Lambda$-term it
may be helpful to consider a general static spacetime with the
metric (in adapted coordinates)
\begin{equation}
ds^2 = \varphi^2 dt^2 + g_{ik}dx^i dx^k,
\end{equation}
where $\varphi$ and $g_{ik}$ depend only on the spatial coordinate
$x^i$. The component $R_{00}$ of the Ricci tensor is given by
$R_{00} = \bar{\Delta}\varphi/ \varphi$, where $\bar{\Delta}$ is
the three-dimensional Laplace operator for the spatial metric $-g_{ik}$ in
(4) (see,e.g., \cite{15}). Let us write Eq. (1) in the form
\begin{equation}
G_{\mu\nu} = \kappa (T_{\mu\nu} + T_{\mu\nu}^{\Lambda})
\quad\quad (\kappa =  8\pi G),
\end{equation}
with
\begin{equation}
T_{\mu\nu}^{\Lambda} = \frac{\Lambda}{8\pi G} g_{\mu\nu}.
\end{equation}
This has the form of the energy-momentum tensor of an ideal fluid,
with energy density $\rho_\Lambda = \Lambda/8\pi G$ and pressure
$p_\Lambda = -\rho_\Lambda$. For an ideal fluid at rest Einstein's
field equation implies
\begin{equation}
\frac{1}{\varphi} \bar{\Delta} \varphi = 4 \pi G \Bigl[ (\rho +
3p) + \underbrace{(\rho_\Lambda + 3p_\Lambda)}_{-2\rho_\Lambda} 
\Bigr].
 \end{equation}
Since the energy density and the pressure appear in the
combination $\rho + 3p$, we understand that a positive
$\rho_\Lambda$ leads to a repulsion (as in (3)). In the Newtonian
limit we have $\varphi \simeq 1 + \phi \; (\phi$ : Newtonian potential)
and $p\ll\rho$, hence we obtain the modified Poisson equation
\begin{equation}
\Delta\phi = 4\pi G(\rho - 2\rho_\Lambda).
\end{equation}
This is the correct Newtonian limit.

As a result of revised values of the Hubble parameter and the
development of the modern theory of stellar evolution in the
1950s, the controversy over ages was resolved and the
$\Lambda$-term became again unnecessary. (Some tension 
remained for values of the Hubble parameter at the higher end of
recent determinations.)

However, in 1967 it was revived again in order to explain why
quasars appeared to have redshifts that concentrated near the
value $z=2$. The idea was that quasars were born in the hesitation
era  \cite{16}. Then quasars at greatly different distances can
have almost the same redshift, because the universe was almost
static during that period. Other arguments in favor of this
interpretation were based on the following peculiarity. When the
redshifts of emission lines in quasar spectra exceed 1.95, then
redshifts of absorption lines in the same spectra were, as a rule,
equal to 1.95. This was then quite understandable, because quasar
light would most likely have crossed intervening galaxies during
the epoch of suspended expansion, which would result in almost
identical redshifts of the absorption lines. However, with more
observational data evidence for the $\Lambda$-term dispersed for
the third time.

Let me conclude this historical review with a few remarks on the
{\it quantum aspect} of the $\Lambda$-problem. Since quantum
physicists had so many other problems, it is not astonishing that
in the early years they did not worry about this subject. An
exception was Pauli, who wondered in the early 1920s whether the
zero-point energy of the radiation field could be gravitationally
effective.

As background I recall that Planck had introduced the zero-point
energy with somewhat strange arguments in 1911. The physical role
of the zero-point energy was much discussed in the days of the old
Bohr-Sommerfeld quantum theory. From Charly Enz and Armin Thellung
-- Pauli's last two assistants -- I have learned that Pauli had
discussed this issue extensively with O.Stern in Hamburg. Stern
had calculated, but never published, the vapor pressure difference
between the isotopes 20 and 22 of Neon (using Debye theory). He
came to the conclusion that without zero-point energy this
difference would be large enough for easy separation of the
isotopes, which is not the case in reality. These considerations
penetrated into Pauli's lectures on statistical mechanics
\cite{17} (which I attended). The theme was taken up in an article
by Enz and Thellung  \cite{18}. This was originally written as a
birthday gift for Pauli, but because of Pauli's early death,
appeared in a memorial volume of Helv.Phys.Acta.

>From Pauli's discussions with Enz and Thellung we know that Pauli
estimated the influence of the zero-point energy of the radiation
field -- cut off at the classical electron radius -- on the radius
of the universe, and came to the conclusion that it ``could not
even reach to the moon''.

When, as a student, I heard about this, I checked Pauli's
unpublished\footnote{A trace of this is in Pauli's Handbuch
article  \cite{19} on wave mechanics in the section where he
discusses the meaning of the zero-point energy of the quantized
radiation field.} remark by doing the following little
calculation:

In units with $\hbar=c=1$ the vacuum energy density of the
radiation field is
\[     <\rho>_{vac} = \frac{8\pi}{(2\pi)^3}\int_0^{\omega_{max}}
    \frac{\omega}{2}\omega^2 d\omega  
             =  \frac{1}{8\pi^2} \omega_{max}^4 , \]
with
\begin{displaymath}
\omega_{max} = \frac{2\pi}{\lambda_{max}} = \frac{2\pi m_e}{\alpha}.
\end{displaymath}
The corresponding radius of the Einstein universe in Eq.(2) would
then be ($M_{pl}\equiv 1/\sqrt{G}$)
\[a = \frac{\alpha^2}{(2\pi)^{\frac{2}{3}}} \frac{M_{pl}}{m_e} \frac{1}{m_e}
\sim 31 km. \]
This is indeed less than the distance to the moon.
(It would be more consistent to use the curvature radius of the
static de Sitter solution; the result is the same, up to the
factor $\sqrt{3}$.)

For decades nobody else seems to have worried about contributions
of quantum fluctuations to the cosmological constant. As far as I know,
Zel'dovich was the first who came back to this issue in two papers
\cite{18a} during the third renaissance period of the $\Lambda$-term, 
but before the advent of spontaneously broken gauge theories. The following
remark by him is particularly interesting. Even if one assumes completely ad hoc
that the zero-point contributions to the vacuum energy density are
exactly cancelled by a bare term (see eq.(29) below), there still remain
higher-order effects. In particular, {\it gravitational} interactions between
the particles in the vacuum fluctuations are expected on dimensional grounds
to lead to a gravitational self-energy density of order $G\mu^6$, where $\mu$
is some cut-off scale. Even for $\mu$ as low as 1 GeV (for no good reason) 
this is about 9 orders of magnitude larger than the observational bound 
(discussed later).

\section{Vacuum fluctuations, vacuum energy}

{\it Without gravity}, we do not care about the absolute energy of
the vacuum, because only energy {\it differences} matter. In
particular, differences of vacuum energies are relevant in many
instances, whenever a system is studied under varying external
conditions. A beautiful example is the {\it Casimir effect}
\cite{20}. In this case the presence of the conducting plates
modifies the vacuum energy density in a manner which depends on
the separation of the plates. This implies an attractive force
between the plates. Precision experiments have recently confirmed
the theoretical prediction to high accuracy (for a recent review,
see \cite{21}). We shall consider other important examples, but
begin with a very simple one which illustrates the main point.

\subsection{A simplified model for the van der Waals force}

Recall first how the zero-point energy of the harmonic oscillator
can be understood on the basis of the canonical commutation relations
$[q,p]= i$. These prevent the {\it simultaneous} vanishing of
the two terms in the Hamiltonian
\begin{equation}
H = \frac{1}{2m} p^2 + \frac{1}{2} m \omega^2 q^2.
\end{equation}
The lowest energy state results from a compromise between the
potential and kinetic energies, which vary oppositely as functions
of the width of the wave function. One understands in this way why
the ground state has an absolute energy which is not zero ({\it
zero-point-energy} $\omega/2$).

Next, we consider two identical harmonic oscillators separated by
a distance $R$, which are harmonically coupled by the dipole-dipole 
interaction energy $\frac{e^2}{R^3} q_1 q_2$. With a simple canonical
transformation we can decouple the two harmonic oscillators and
find for the frequencies of the decoupled ones \begin{math}
\omega_i^2 = \omega^2 \pm \frac{e^2}{m} \frac{1}{R^3}\end{math},
and thus for the ground state energy
\begin{displaymath}
E_0(R) = \frac{1}{2} (\omega_1 + \omega_2) \approx  \omega -
\frac{e^4}{8 \omega^3 R^6 }.
\end{displaymath}
The second term on the right depends on R and gives the van der
Waals force (which vanishes for $\hbar\rightarrow 0$).

\subsection{Vacuum fluctuations for the free radiation field}

Similar phenomena arise for quantized fields. We consider, as a
simple example, the free quantized electromagnetic field
$F_{\mu\nu}(x)$. For this we have for the equal times commutators
the following nontrivial one (Jordan and Pauli \cite {22}):
\begin{equation}
\Bigl[ E_i(\mbox{\boldmath $x$}), B_{jk}(\mbox{\boldmath $x'$})
\Bigr] = i \Bigl(\delta_{ij} \frac{\partial}{\partial x_k}
-\delta_{ik} \frac{\partial}{\partial x_j} \Bigr)
\delta^{(3)}(\mbox{\boldmath $x$} -\mbox{\boldmath $x'$})
\end{equation}
(all other equal time commutators vanish); here $B_{12}=B_{3}$,
and cyclic. This basic commutation relation prevents the
simultaneous vanishing of the electric and magnetic energies. It
follows that the ground state of the quantum field (the vacuum)
has a non-zero absolute energy, and that the variances of
$\mbox{\boldmath$E$}$ and $\mbox{\boldmath$B$}$ in this state are
nonzero. This is, of course, a quantum effect.

In the Schr\"odinger picture the electric field operator has the
expansion
\begin{equation}
\mbox{\boldmath$E$}(\mbox{\boldmath$x$}) =
\frac{1}{(2\pi)^{3/2}}\int\frac{d^3k} {\sqrt{2\omega(k)}}
\sum_{\lambda} \Bigl[i\omega(k)a(\mbox{\boldmath$k$},\lambda)
\mbox{\boldmath$\epsilon $}(\mbox{\boldmath$k$},\lambda)
\exp(i\mbox{\boldmath$k$}\cdot \mbox{\boldmath$x$)} + h.c. \Bigr].
\end{equation}
(We use Heaviside units and always set $\hbar=c=1$.)\\
Clearly,
\[ <\mbox{\boldmath$E$}(\mbox{\boldmath$x$})>_{vac} = 0. \]
The expression
$<\mbox{\boldmath$E$}^2(\mbox{\boldmath$x$})>_{vac}$ is not
meaningful. We smear $\mbox{\boldmath$E$}(\mbox{\boldmath$x$})$
with a real test function $f$:
\begin{eqnarray*}
\mbox{\boldmath$E$}_f(\mbox{\boldmath$x$})
&=&\int \mbox{\boldmath$E$}
(\mbox{\boldmath$x$}+\mbox{\boldmath$x'$}) f(\mbox{\boldmath$x'$}) d^3 x' \\
&=& \frac{1}{(2\pi)^{3/2}} \int \frac{d^3k}{\sqrt{2\omega(k)}} \sum_{\lambda}
\Bigl[i\omega(k)a(\mbox{\boldmath$k$},\lambda)
\mbox{\boldmath$\epsilon$}(\mbox{\boldmath$k$},\lambda)
 \exp(i\mbox{\boldmath$k$}\cdot
\mbox{\boldmath$x$})\hat{f}(\mbox{\boldmath$k$})
  +  h.c.  \Bigr], \\
\end{eqnarray*}
where

\[   \hat{f}(\mbox{\boldmath$k$}) = \int f(\mbox{\boldmath$x$})
\exp(i\mbox{\boldmath$k\cdot x$})\, d^3x. \]
It follows immediately that
\[ <\mbox{\boldmath$E$}_f^2(\mbox{\boldmath$x$})> = 2\int\frac{d^3k}{(2\pi)^3}
\frac{\omega}{2}|\hat{f}(\mbox{\boldmath$k$})|^2. \]
For a sharp momentum cutoff $\hat{f}(\mbox{\boldmath$k$}) =
\Theta(\mathcal{K} -|\mbox{\boldmath$k$}|)$,
we have
\begin{equation}
<\mbox{\boldmath$E$}_f^2(x)>_{vac} =
\frac{1}{2\pi^2}\int_0^\mathcal{K}\omega^3 d\omega =
\frac{\mathcal{K}^4}{8\pi^2}. \] The vacuum energy density for
$|k|\le{\mathcal {K}}$ is
\[ \rho_{vac} =  \frac{1}{2}<\mbox{\boldmath$E$}^2 + \mbox{\boldmath$B$}^2>_{vac}
 = <\mbox{\boldmath$E$}^2>_{vac}
 = \frac{\mathcal{K}^4}{8\pi^2}.
 \end{equation}
Again, without gravity we do not care, but as in the example
above, this vacuum energy density becomes interesting when we
consider varying external conditions. This leads us to the next
example.

\subsection{The Casimir effect}

This well-known instructive example has already been mentioned.
Let us consider the simple configuration of two large parallel
perfectly conducting plates, separated by the distance $d$. The
vacuum energy per unit surface of the conductor is, of course,
divergent and we have to introduce some intermediate
regularization. Then we must subtract the free value (without the
plates) for the same volume. Removing the regularization
afterwards, we end up with a finite $d$-dependent result.

Let me give for this simple example the details for two different
regularization schemes. If the plates are parallel to the
$(x_1,x_2)$-plane, the vacuum energy per unit surface is
(formally):
\begin{equation}
\mathcal{E}_{vac} = \sum_{l=0}^\infty\int_{\mathbf{R}^2}\Bigl[
k_1^2 + k_2^2 + (\frac{l\pi}{d})^2 \Bigr ]^{1/2}
\frac{d^2k}{(2\pi)^2}.
\end{equation}
In the first regularization we replace the frequencies $\omega$ of
the allowed modes (the square roots in Eq.(13)) by $\omega\exp(-
\frac{\alpha}{\pi}\omega)$, with a parameter $\alpha$. A polar
integration can immediately be done, and we obtain (leaving out the
$l=0$ term, which does not contribute after subtraction of the
free case):
\begin{eqnarray*}
\mathcal{E}_{vac}^{reg}& = &
\frac{\pi^2}{4}\sum_{l=1}^\infty(\frac{l}{d})^3
\int_0^\infty \exp(-\alpha\frac{l}{d}\sqrt{1+z})\sqrt{1+z}dz \\
& = & -\frac{\pi^2}{4} \frac{\partial^3}{\partial\alpha^3}\sum_{l=1}^\infty
\int_0^\infty \exp(-\alpha\frac{l}{d}\sqrt{1+z})\frac{dz}{1+z}.
 \end{eqnarray*}
In the last expression the sum is just a geometrical series. After
carrying out one differentiation the integral can easily been
done, with the result
\begin{equation}
\mathcal{E}_{vac}^{reg} = \frac{\pi^2}{2d} \frac{\partial^2}{\partial\alpha^2}
\frac{d/\alpha}{e^{\alpha/d}-1}.
\end{equation}
Here we use the well-known formula
\begin{equation}
\frac{x}{e^x-1} = \sum_{n=0}^\infty \frac{B_n}{n!} x^n ,
\end{equation}
where the $B_n$ are the Bernoulli numbers. It is then easy to
perform the renormalization (subtraction of the free case).
Removing afterwards the regularization ($\alpha\rightarrow 0$)
gives the renormalized result:
\begin{equation}
\mathcal{E}_{vac}^{ren} = -\frac{\pi^2}{d^3} \frac{B_4}{4!} =
-\frac{\pi^2}{720}\frac{1}{d^3}.
\end{equation}
The corresponding force per unit area is
\begin{equation}
\mathcal{F} = -\frac{\pi^2}{240}  \frac{1}{d^4} =
-\frac{0.013}{(d(\mu m))^4} {dyn/cm^2}.
\end{equation}

Next, I describe the $\zeta$-function regularization. This method
has found many applications in quantum field theory, and is
particularly simple in the present example.

Let me first recall the definition of the $\zeta$-function
belonging to a selfadjoint operator $A$ with a purely discrete
spectrum, $ A = \sum_n \lambda_n P_n$, where the $\lambda_n$ are
the eigenvalues and $P_n$ the projectors on their eigenspaces with
dimension $g_n$. By definition
\begin{equation}
\zeta_{A}(s) = \sum_n \frac{g_n}{\lambda_n^s}.
\end{equation}
Assume that $A$ is positive and that the trace of $A^\frac{1}{2}$
exists, then
\begin{equation}
Tr A^\frac{1}{2} = \zeta_{A}(-1/2).
\end{equation}
Formally, the sum (13) is -- up to a factor 2 -- the trace (19)
for $A = -\Delta $, where $\Delta$ is the Laplace operator for the
region between the two plates with the boundary conditions imposed
by the ideally conducting plates. (Recall that the term with $l=0$
is irrelevant.) Since this trace does not exist, we proceed as
follows ($\zeta$- function regularization): Use that
$\zeta_{A}(s)$ is well-defined for $\Re s>2$ and that it can
analytically be continued to some region with $\Re s<2$ including
$s=-1/2$ (see below), we can {\it define} the regularized trace by
Eq.(19).

The short calculation involves the following steps. For $s>2$ we
have
\begin{eqnarray}
\zeta_{-\Delta}(s) & = & 2\sum_{l=1}^{ \infty}\int_{\textbf{R}^2 }
\frac{1}{\Bigl[ k_1^2 + k_2^2 +
(\frac{l\pi}{d})^2 \Bigr] ^s }\frac{d^2k}{(2\pi)^2} \\ & = &  \frac{1}{2\pi}
\frac{\Gamma(s-1)}{\Gamma(s)}\zeta_{R}(2s-2) (\frac{\pi}{d})^{2(1-s)} ,
\end{eqnarray}
where $\zeta_{R}(s)$ is the $\zeta$-function of Riemann. For the
analytic continuation we make use of the well-known formula
\begin{equation}
\zeta_{R}(1-s) = \frac{1}{(2\pi)^s}2\Gamma(s)\cos(\frac{\pi s}{2}) \zeta_{R}(s)
\end{equation}
and find
\begin{equation}
\zeta_{-\Delta}(-1/2) = -\frac{\pi^2}{360}\frac{1}{d^3}.
\end{equation}
This gives the result (16).

For a mathematician this must look like black magic, but that's
the kind of things physicists are doing to extract physically
relevant results from mathematically ill-defined formalisms.

One can similarly work out the other components of the
energy-momentum tensor, with the result
\begin{equation}
<T^{\mu\nu}>_{vac} = \frac{\pi^2}{720}\frac{1}{d^4}diag(-1,1,1,-3).
\end{equation}
This can actually be obtained without doing additional
calculations, by using obvious symmetries and general properties of
the energy-momentum tensor.

By now the literature related to the Casimir effect is enormous.
For further information we refer to the recent book \cite {23}.

\subsection{Radiative corrections to Maxwell's equations}

Another very interesting example of a vacuum energy effect was
first discussed by Heisenberg and Euler \cite{24} , and later by
Weisskopf \cite{25}.

When quantizing the electron-positron field one also encounters an
infinite vacuum energy ( the energy of the Dirac sea):
\[ \mathcal{E}_0 = -\sum_{\mbox{\boldmath$p$},\sigma}
 \varepsilon_{\mbox{\boldmath$p$},\sigma} ^{(-)}, \]
where $-\varepsilon^{(-)}_{\mbox{\boldmath$p$},\sigma}$ are the
negative frequencies of the solutions the Dirac equation. Note that
$\mathcal{E}_0$ is {\it negative}, which  already early gave rise
to the hope that perhaps fermionic and bosonic contributions
might compensate. Later, we learned that this indeed happens in
theories with unbroken supersymmetries. The constant
$\mathcal{E}_o$ itself again has no physical meaning. However, if
an external electromagnetic field is present, the energy levels
$\varepsilon_{ \mbox{\boldmath$p$},\sigma}^{(-)}$ will change.
These changes are finite and {\it physically significant}, in that
they alter the equations for the electromagnetic field in vacuum.

The main steps which lead to the correction $\mathcal{L}'$ of
Maxwell's Lagrangian $\mathcal{L}_o =
-\frac{1}{4}F_{\mu\nu}F^{\mu\nu} $ are the following ones (for
details see \cite{26}):

First one shows (Weisskopf) that
\[ \mathcal{L}' = -\Bigl[ \mathcal{E}_0 - \mathcal{E}_0|_{E=B=0} \Bigr]. \] 
After a charge renormalization, which ensures that $\mathcal{L}'$
has no quadratic terms, one arrives at a finite correction. For
almost homogeneous fields it is a function of the invariants
\begin{eqnarray}
\mathcal{F} &=& \frac{1}{4}F_{\mu\nu}F^{\mu\nu} = \frac{1}{2}
(\mbox{\boldmath$B$}^2 -
\mbox{\boldmath$E$}^2),\\
\mathcal{G}^2 &=& \Bigl( \frac{1}{4}F_{\mu\nu}^{\ast}F^{\mu\nu}\Bigr)^2 =
(\mbox{\boldmath$E$}\cdot \mbox{\boldmath$B$})^2.
\end{eqnarray}
In \cite{26} this function is given in terms of a 1-dimensional
integral. For weak fields one finds
\begin{equation}
\mathcal{L}' = \frac{2\alpha^2}{45m^4}\Bigl[ (\mbox{\boldmath$E$}^2 -
 \mbox{\boldmath$B$}^2)^2  +
7(\mbox{\boldmath$E\cdot B$})^2 \Bigr] +\cdot\cdot\cdot  .
\end{equation}

An alternative efficient method to derive this result again makes
use of the $\zeta$-function regularization (see, e.g., \cite{27}).

The correction (27) gives rise to electric and magnetic polarization
vectors of the vacuum. In particular, the refraction index for light
propagating perpendicular to a static homogeneous magnetic field depends
on the polarization direction. This is the vacuum analog of the well-known 
Cotton-Mouton effect in optics. As a result, an initially linearly polarized
light beam becomes elliptic. In spite of great efforts it has not yet been
possible to observe this effect.\footnote{After my talk Carlo Rizzo informed
me about two current projects to measure the Cotton-Mouton effect in vacuum \cite{
{27a}}.}

For other fluctuation-induced forces, in particular in condensed
matter physics, I refer to the review article \cite{28}.

\section{ Vacuum energy and gravity}

When we consider the coupling to gravity, the vacuum energy
density acts like a cosmological constant. In order to see this,
first consider the vacuum expectation value of the energy-momentum
tensor in Minkowski spacetime. Since the vacuum state is Lorentz
invariant, this expectation value is an invariant symmetric
tensor, hence proportional to the metric tensor. For a curved
metric this is still the case, up to higher curvature terms:
\begin{equation}
<T_{\mu\nu}>_{vac} = g_{\mu\nu}\rho_{vac} + \textit{higher curvature terms}.
\end{equation}
The {\it effective} cosmological constant, which controls the
large scale behavior of the Universe, is given by
\begin{equation}
\Lambda = 8\pi G\rho_{vac}+\Lambda_{0},
\end{equation}
where $\Lambda_0$ is a bare cosmological constant in Einstein's
field equations.

We know from astronomical observations discussed later in Sect. 5 and 6 that
$\rho_\Lambda\equiv\Lambda/8\pi G$ can
not be larger than about the critical density:
\begin{eqnarray}
\rho_{crit} &=& \frac{3H_0^2}{8\pi G} \nonumber \\
&=& 1.88\times 10^{-29} h_0^2 \textit {g}\textit {cm}^{-3} \\
&=& 8\times10^{-47} h_0^2 \textit{GeV}^{4}, \nonumber
\end{eqnarray}
where $h_0$ is the {\it reduced Hubble parameter}
\begin{equation}
h_0 = H_0/(100\textit{km}\textit{s}^{-1}\textit{Mpc}^{-1})
\end{equation}
and is close to 0.6 \cite{29}.

It is a complete mystery as to why the two terms in (29) should
almost exactly cancel. This is -- more precisely stated -- the
famous $\Lambda$-problem. It is true that we are unable to
calculate the vacuum energy density in quantum field theories,
like the Standard Model of particle physics. But we can attempt to
make what appear to be  reasonable order-of-magnitude estimates
for the various contributions. This I shall describe in the
remainder of this section. The expectations will turn out to be in
gigantic conflict with the facts.

\subsection*{Simple estimates of vacuum energy contributions}

If we take into account the contributions to the vacuum energy
from vacuum fluctuations in the fields of the Standard Model up to
the currently explored energy, i.e., about the electroweak scale
$M_F = G_F^{-1/2}\approx 300 GeV  (G_F:$ Fermi coupling constant), 
we cannot expect an almost complete
cancellation, because there is {\it no symmetry principle} in this
energy range that could require this. The only symmetry principle
which would imply this is {\it supersymmetry}, but supersymmetry
is broken (if it is realized in nature). Hence we can at best
expect a very imperfect cancellation below the electroweak scale,
leaving a contribution of the order of $M_F^4$ . (The
contributions at higher energies may largely cancel if
supersymmetry holds in the real world.)

We would reasonably expect that the vacuum energy density is at
least as large as the condensation energy density of the QCD phase
transition to the broken phase of chiral symmetry. Already this is
far too large: $ \sim \Lambda_{QCD}^4/ 16\pi^2  \sim 10^{-4}
\textit{GeV}^4$; this is {\it more than 40 orders of magnitude
larger} than $\rho_{crit}$. Beside the formation of quark
condensates $<\bar{q}q>$ in the QCD vacuum which break chirality,
one also expects a gluon condensate $<G^{\mu\nu}_{a}
G_{a\mu\nu}>\\* \sim \Lambda_{QCD}^4.$ This produces a significant
vacuum energy density as a result of a dilatation anomaly: If
$\Theta_\mu^\mu$ denotes the ``classical'' trace of the
energy-momentum tensor, we have \cite{30}
\begin{equation}
T^\mu_\mu = \Theta^\mu_\mu + \frac{\beta(g_3)}{2g_3}
G^{\mu\nu}_{a}G_{a\mu\nu},
\end{equation}
where the second term is the QCD piece of the trace anomaly
($\beta(g_3)$ is the $\beta$-function of QCD that determines the
running of the strong coupling constant). I recall that this
arises because a scale transformation is no more a symmetry if
quantum corrections are included. Taking the vacuum expectation
value of (32), we would again naively expect that $<\Theta_\mu^
\mu>$ is of the order $M_F^4$. Even if this should vanish for some
unknown reason, the anomalous piece is cosmologically gigantic.
The expectation value $<G^{\mu\nu}_a G_{a\mu\nu}>$ can be
estimated with QCD sum rules \cite{31}, and gives

\begin{equation}
<T^{\mu}_\mu>^{anom}\sim (350 MeV)^4,
\end{equation}
about 45 orders of magnitude larger than $\rho_{crit}$. This
reasoning should show convincingly that the cosmological constant
problem is indeed a profound one. (Note that there is some analogy
with the (much milder) strong CP problem of QCD. However, in
contrast to the $\Lambda$-problem, Peccei and Quinn \cite{32} have
shown that in this case there is a way to resolve the conundrum.)

Let us also have a look at the Higgs condensate of the electroweak
theory. Recall that in the Standard Model we have for the Higgs
doublet $\Phi$ in the broken phase for
$<\Phi^{*}\Phi>\equiv\frac{1}{2}\phi^2$ the potential
\begin{equation}
V(\phi) = -\frac{1}{2}m^2\phi^2 + \frac{\lambda}{8}\phi^4.
\end{equation}
Setting as usual $\phi=v+H$, where $v$ is the value of $\phi$ where $V$ has
its minimum,
\begin{equation}
v= \sqrt{\frac{2m^2}{\lambda}} = 2^{-1/4}G_F^{-1/2}  \sim 246 GeV,
\end{equation} we find that the Higgs mass is related to $\lambda$ by
$\lambda= M_{H}^2/v^2$. For  $\phi=v$
we obtain the energy density of the Higgs condensate
\begin{equation}
V(\phi=v)=-\frac{m^4}{2\lambda}= -\frac{1}{8\sqrt{2}}M_F^2M_H^2=
\mathcal{O}(M_F^4).
\end{equation}
We can, of course, add a constant $V_0$ to the potential (34) such
that it cancels the Higgs vacuum energy in the broken phase --
including higher order corrections. This again requires an extreme
fine tuning. A remainder of only $\mathcal{O}(m_e^4)$, say, would
be catastrophic. This remark is also highly relevant for models of
inflation and quintessence.

In attempts beyond the Standard Model the vacuum energy problem so
far remains, and often becomes even worse. For instance, in supergravity
theories with spontaneously broken supersymmetry there is the following
simple relation between the gravitino mass $m_g$ and the vacuum energy 
density 
\[ \rho_{vac} = \frac{3}{8\pi G}m_g^2.\]
Comparing this with eq.(30) we find 
\[ \frac{\rho_{vac}}{\rho_{crit}} \simeq 10^{122}\Bigl (\frac{m_g}{m_{Pl}}\Bigr )^2.\]
Even for $m_g \sim 1\; eV$ this ratio becomes $10^{66}$. ($m_g$ is related to 
the parameter $F$ characterizing the strength of the supersymmetry breaking
by $m_g = (4\pi G/3)^{1/2} F$, so $m_g\sim 1\; eV$ corresponds to $F^{1/2} \sim
100\; TeV$.)

Also string theory has not yet offered convincing clues why the cosmological 
constant is so extremely small. The main reason is that a {\it low energy
mechanism} is required, and since supersymmetry is broken, one again expects 
a magnitude of order $M_F^4$, which is {\it at least 50 orders of magnitude too large}
(see also \cite{33}). However, non-supersymmetric physics in string theory
is at the very beginning and workers in the field hope that further progress
might eventually lead to an understanding of the cosmological constant problem.

I hope I have convinced you, that there is something profound that
we do not understand at all, certainly not in quantum field
theory, but so far also not in string theory. ( For other recent
reviews, see also \cite{34}, \cite{34a}, and \cite{35}. These contain more
extended lists of references.)

This is the moment to turn to the astronomical and astrophysical
aspects of our theme. Here, exciting progress can be reported.

\section{Luminosity-redshift relation for type Ia supernovas}
A few years ago the Hubble diagram for type Ia supernovas gave the
first serious evidence for an accelerating Universe. Before
presenting and discussing these exciting results we recall some
theoretical background.

\subsection{Theoretical redshift-luminosity relation}
In cosmology several different distance measures are in use. They
are all related by simple redshift factors. The one which is
relevant in this Section is the {\it luminosity distance} $D_L$,
defined by
\begin{equation}
D_L = (\mathcal{L}/4\pi\mathcal{F})^{1/2},
\end{equation}
where $\mathcal{L}$ is the intrinsic luminosity of the source and
$\mathcal{F}$ the observed flux.

We want to express this in terms of the redshift $z$ of the source
and some of the cosmological parameters. If the comoving radial
coordinate $r$ is chosen such that the Friedmann- Lema\^{\i}tre
metric takes the form
\begin{equation}
g = dt^2 - a^2(t) \Bigl[ \frac{dr^2}{1-kr^2} + r^2d\Omega^2 \Bigr ],\;\;\;  k = 0, \pm1,
\end{equation}
then we have
\[\mathcal{F}dt_0 = \mathcal{L}dt_e \cdot\frac{1}{1+z}\cdot
\frac{1}{4\pi(r_e a(t_0))^2}. \]

The second factor on the right is due to the redshift of the
photon energy; the indices $0, e$ refer to the present and
emission times, respectively. Using also $1+z=a(t_0)/ a(t_e)$, we
find in a first step:
\begin{equation}
D_L(z) = a_0 (1+z)r(z) \;\;\; (a_0\equiv a(t_0)).
\end{equation}

We need the function $r(z)$. From \[ dz=
-\frac{a_0}{a}\frac{\dot{a}}{a}dt,\;\;\;\; dt =
-a(t)\frac{dr}{\sqrt{1-kr^2}} \] for light rays, we see that
\begin{equation}
\frac{dr}{\sqrt{1-kr^2}} = \frac{1}{a_0} \frac{dz}{H(z)}  \;\;\;\;
( H(z)=\frac{\dot{a}}{a}).
\end{equation}
Now, we make use of the Friedmann equation
\begin{equation}
H^2 + \frac{k}{a^2} = \frac{8\pi G}{3}\rho .
\end{equation}
Let us  decompose the total energy-mass density $\rho $ into
nonrelativistic (NR), relativistic (R), $\Lambda$,
quintessence (Q), and possibly other contributions
\begin{equation}
\rho = \rho_{NR} + \rho_R + \rho_\Lambda + \rho_Q + \cdots .
\end{equation}
For the relevant cosmic period we can assume that the ``energy equation''
\begin{equation}
\frac{d}{da}(\rho a^3) = -3pa^2
\end{equation}
also holds for the individual components $X=NR,R,\Lambda,Q,\cdots$. If $w_X
\equiv p_X/\rho_X $ is constant,this implies that
\begin{equation}
\rho_Xa^{3(1+w_X)} = const.
\end{equation}
Therefore,
\begin{equation}
 \rho = \sum_X (\rho_X a^{3(1+w_X)})_0 \frac{1}{a^{3(1+w_X)}} =
 \sum_X (\rho_X)_0 (1+z)^{3(1+w_X)}.
 \end{equation}
Hence the Friedmann equation (41) can be written as
\begin{equation}
\frac{H^2(z)}{H_0^2} + \frac{k}{H_0^2 a_0^2 }(1+z)^2 =
\sum_X \Omega_X (1+z)^{3(1+w_X)},
\end{equation}
where $\Omega_X$ is the dimensionless density parameter for the species $X$,
\begin{equation}
\Omega_X = \frac{(\rho_X)_0}{\rho_{crit}}.
\end{equation}
Using also the curvature parameter $\Omega_K\equiv-k/H_0^2 a_0^2$,
we obtain the useful form\begin{equation}
 H^2(z) = H_0^2 E^2(z;\Omega_K,\Omega_X),
\end{equation}
with
\begin{equation}
E^2(z;\Omega_K,\Omega_X) = \Omega _K(1+z)^2 + \sum_X \Omega_X (1+z)^{3(1+w_X)}.
\end{equation}
Especially for $z=0$ this gives
\begin{equation}
\Omega_K + \Omega_0 = 1,\;\;\;\; \Omega_0 \equiv \sum_X \Omega_X.
\end{equation}
If we use (48) in (40), we get
\begin{equation}
\int_0^{r(z)} \frac{dr}{\sqrt{1-r^2}} = \frac{1}{H_0 a_0}\int_0^z \frac{dz'}{E(z')}
\end{equation}
and thus
\begin{equation}
r(z) = \mathcal{S}(\chi(z)),
\end{equation}
where
\begin{equation}
\chi(z) = |\Omega_K|^{1/2} \int_0^z \frac{dz'}{E(z')}
\end{equation}
and
\begin{equation}
\mathcal{S} = \left\{ 
\begin{array}{ll}
\sin \chi & :  k=1  \\ 
\chi & :  k=0 \\
\sinh \chi & :  k=1  
\end{array} \right. 
\end{equation} 
Inserting this in (39) gives finally the relation we were looking for
\begin{equation}
D_L(z) = \frac{1}{H_0}\mathcal{D}_L(z;\Omega_K,\Omega_X) ,
\end{equation}
with
\begin{equation}
\mathcal{D}_L(z;\Omega_K,\Omega_X) = (1+z)\frac{1}{|\Omega_K|^{1/2}}\mathcal{S}
(|\Omega_K|^{1/2}\int_0^z \frac{dz'}{E(z')}).
\end{equation}
Note that for a flat universe, $ \Omega_K =0$ or equivalently  $\Omega_0 =1$, 
the ``Hubble-constant-free''luminosity distance is
\begin{equation}
\mathcal{D}_L(z) = (1+z)\int_0^z\frac{dz'}{E(z')}.
\end{equation}

Astronomers use as logarithmic measures of $ \mathcal{L}$ and 
$\mathcal{F}$ the {\it absolute and apparent  magnitudes}
\footnote{Beside the (bolometric) magnitudes $m,M$, astronomers 
also use magnitudes $m_B,m_V,\ldots$ referring to certain 
wavelength bands $B$ (blue), $V$ (visual), and so on.}, denoted by 
$M$ and $m$, respectively. The conventions are chosen such that 
the {\it distance modulus} $m-M$ is related to $D_L$ as follows 
\begin{equation}
m-M = 5 \log \Bigl( \frac{D_L}{1 Mpc}\Bigr) + 25.
\end{equation}
Inserting the representation (55), we obtain the following
relation between the apparent magnitude $m$ and the redshift $z$:
\begin{equation}
m= \mathcal{M} + 5 \log \mathcal{D}_L(z;\Omega_K,\Omega_X),
\end{equation}
where, for our purpose, $ \mathcal{M}=M-5\log H_0 -25$ is an
uninteresting fit parameter. The comparison of this theoretical
{\it magnitude redshift relation} with data will lead to
interesting restrictions for the cosmological 
$\Omega$-parameters. In practice often only $\Omega_M$ and 
$\Omega_\Lambda$ are kept as independent parameters, where from now
on the subscript $M$ denotes (as in most papers) nonrelativistic
matter.

The following remark about {\it degeneracy curves} in the
$\Omega$-plane is important in this context. For a fixed $z$ in
the presently explored interval, the contours defined by the
equations $\mathcal{D}_L(z;\Omega_M,\Omega_\Lambda) = const $ have
little curvature, and thus we can associate an approximate slope
to them. For $z=0.4$ the slope is about 1 and increases to 1.5-2
by $z=0.8$ over the interesting range of $\Omega_M $ and
$\Omega_\Lambda $. Hence even quite accurate data can at best
select a strip in the $\Omega$-plane, with a slope in the range
just discussed. This is the reason behind the shape of the
likelihood regions shown later (Fig.2).

In this context it is also interesting to determine the dependence of the
{\it deceleration parameter}
\begin{equation}
q_0 = - \Bigl( \frac{a\ddot{a}}{\dot{a}^2} \Bigr )_0
\end{equation}
on $\Omega_M$ and $\Omega_\Lambda$.
At an any cosmic time we obtain from (3) and (45)
\begin{equation}
-\frac{\ddot{a}a}{\dot{a}^2} = \frac{1}{2}\frac{1}{E^2(z)}\sum_X
\Omega_X (1+z)^{3(1+w_X)}(1+3w_X).
\end{equation}
For $z=0$ this gives
\begin{equation}
q_0 = \frac{1}{2}\sum_X \Omega_X (1+3w_X) = \frac{1}{2} (\Omega_M
- 2\Omega_\Lambda + \cdots).
\end{equation}
The line $q_0=0 \; (\Omega_\Lambda = \Omega_M /2 )$ separates
decelerating from accelerating universes at the present time. For
given values of $\Omega_M, \Omega_\Lambda$, etc, (61) vanishes for
$z$ determined by
\begin{equation}
\Omega_M(1+z)^3 - 2\Omega_\Lambda +\cdots = 0.
\end{equation}
This equation gives the redshift at which the deceleration period
ends (coasting redshift).

\subsection{Type Ia supernovas as standard candles}

It has long been recognized that supernovas of type Ia are
excellent standard candles and are visible to cosmic distances
\cite{36} (the record is at present at a redshift of about 1.7).
At relatively closed distances they can be used to measure the
Hubble constant, by calibrating the absolute magnitude of nearby
supernovas with various distance determinations (e.g., Cepheids).
There is still some dispute over these calibration resulting in
differences of about 10\% for $H_0$. (For a review see, e.g.,
\cite{29}.)

In 1979 Tammann \cite{37}and Colgate \cite{38} independently
suggested that at higher redshifts this subclass of supernovas can
be used to determine also the deceleration parameter. In recent
years this program became feasible thanks to the development of
new technologies which made it possible to obtain digital images
of faint objects over sizable angular scales, and by making use of
big telescopes such as Hubble and Keck.

There are two major teams investigating high-redshift SNe Ia,
namely the `Supernova Cosmology Project' (SCP) and the `High-Z
Supernova search Team' (HZT). Each team has found a large number
of SNe, and both groups have published almost identical results.
(For up-to-date information, see the home pages \cite{39} and
\cite{40}.)

Before discussing these, a few remarks about the nature and
properties of type Ia SNe should be made. Observationally, they
are characterized by the absence of hydrogen in their spectra, and
the presence of some strong silicon lines near maximum. The
immediate progenitors are most probably carbon-oxygen white dwarfs
in close binary systems, but it must be said that these have not
yet been clearly identified. \footnote {This is perhaps not so
astonishing, because the progenitors are presumably faint compact
dwarf stars.}

In the standard scenario a white dwarf accretes matter from a
nondegenerate companion until it approaches the critical
Chandrasekhar mass and ignites carbon burning deep in its interior
of highly degenerate matter. This is followed by  an
outward-propagating nuclear flame leading to a total disruption of
the white dwarf. Within a few seconds the star is converted
largely into nickel and iron. The dispersed nickel radioactively
decays to cobalt and then to iron in a few hundred days. A lot of
effort has been invested to simulate these complicated processes.
Clearly, the physics of thermonuclear runaway burning in
degenerate matter is complex. In particular, since the
thermonuclear combustion is highly turbulent, multidimensional
simulations are required. This is an important subject of current
research. (One gets a good impression of the present status from
several articles in \cite{41}. See also the recent review
\cite{42}.) The theoretical uncertainties are such that, for
instance, predictions for possible evolutionary changes are not
reliable.

It is conceivable that in some cases a type Ia supernova is the
result of a merging of two carbon-oxygen-rich white dwarfs with a
combined mass surpassing the Chandrasekhar limit. Theoretical
modelling indicates, however, that such a merging would lead to a
collapse, rather than a SN Ia explosion. But this issue is still
debated.

In view of the complex physics involved, it is not astonishing
that type Ia  supernovas are not perfect standard candles. Their
peak absolute magnitudes have a dispersion of 0.3-0.5 mag,
depending on the sample. Astronomers have, however learned in
recent years to reduce this dispersion by making use of empirical
correlations between the absolute peak luminosity and light curve
shapes. Examination of nearby SNe showed that the peak brightness
is correlated with the time scale of their brightening and fading:
slow decliners tend to be brighter than rapid ones. There are
also some correlations with spectral properties. Using these
correlations it became possible to reduce the remaining intrinsic
dispersion to $\simeq 0.17 mag$. (For the various methods in use,
and how they compare, see \cite{43}, and references therein.)
Other corrections, such as Galactic extinction, have been applied,
resulting for each supernova in a corrected (rest-frame)
magnitude. The redshift  dependence of this quantity is compared
with the theoretical expectation given by Eqs.(59) and (56).

\subsection{Results}

In Fig.1 the Hubble diagram for the high-redshift supernovas,
published by the SCP and HZT teams \cite{44}, \cite{45}, \cite{46}
is shown. All data have been normalized by the same ($\Delta
m_{15}$) method \cite{47}. In both panels the magnitude
differences relative to an empty universe are plotted. The upper panel shows                
the data for both teams separately. These can roughly be summarized by 
the statement that distant supernovas are in the average {\it about 0.20 magnitudes
fainter than in an empty Friedmann universe}. In the lower panel the data are
redshift binned, and the result for the very distant SN 1999ff at $z\simeq 1.7$
is also shown.

\begin{figure}
\begin{center}
\includegraphics[height=0.7\textheight]{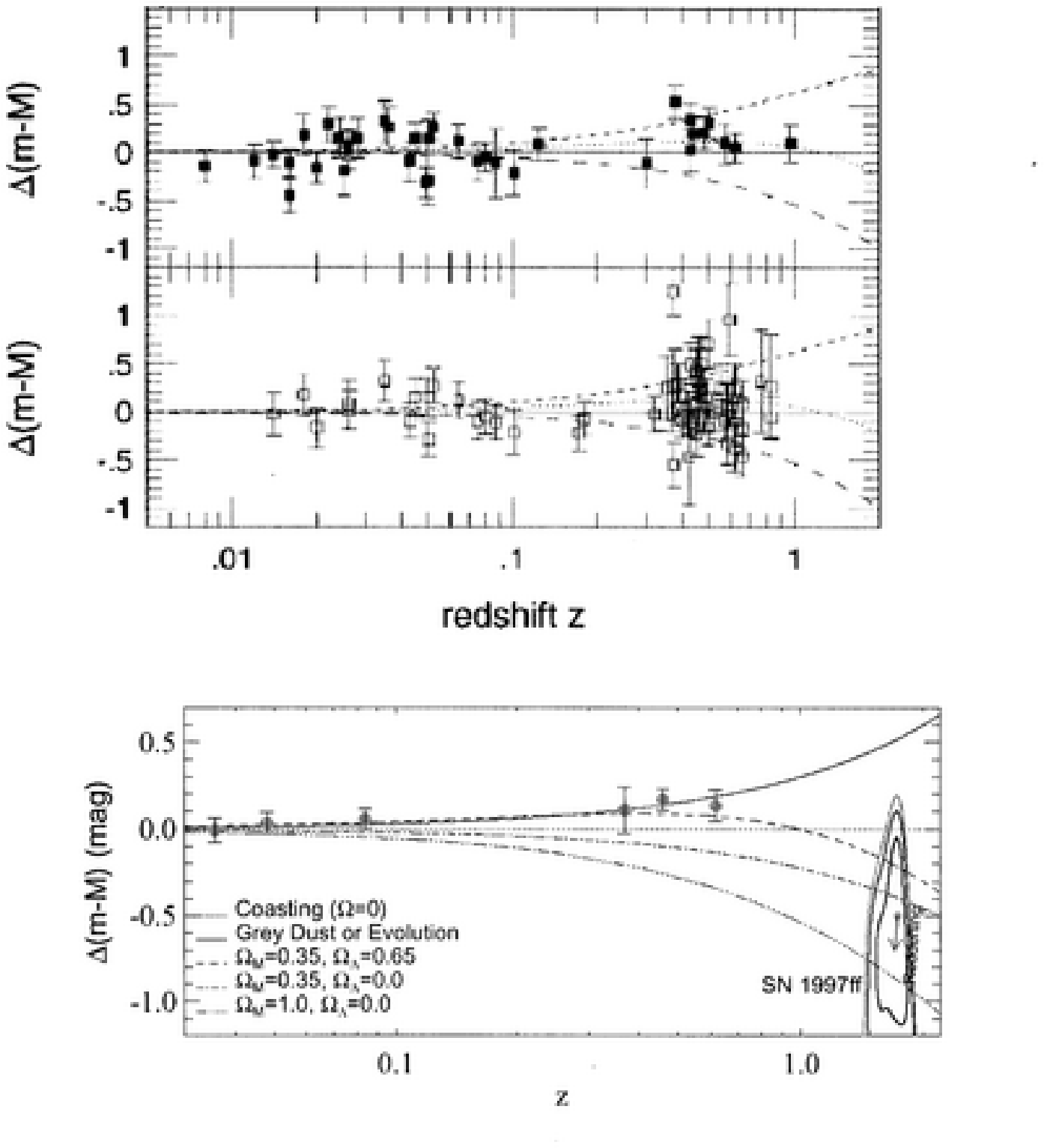}
\caption{Hubble diagram of Type Ia Supernovas minus an 
empty ($\Omega_0$) Universe compared to cosmological models. All
data in the upper panel have been normalized with the same $(\Delta m_{15})$
method (Leibundgut \cite{47}). The filled squares are the data from HZT
\cite{45}, and those of SCP \cite{44} are shown as open squares. 
The parameters
$(\Omega_M,\Omega_\Lambda)$ of the cosmological models are: (1,0) 
(long dashes), (0,1) (dashed line), (0.3,0.7) (dotted line). 
In the lower panel the points are redshift-binned data from both teams 
\cite{49}. A typical curve for grey dust evolution is also shown. 
In spite of the large uncertainties of SN 1999ff at 
$z\simeq 1.7$, simple grey dust evolution seems to be excluded.}
\label{Fig-1}
\end{center}
\end{figure}

The main result of the analysis is presented in Fig.2. Keeping
only $\Omega_M$ and $\Omega_\Lambda$ in Eq.(56) ( whence $\Omega_K
= 1-\Omega_M -\Omega_\Lambda)$ in the fit to the data of 79 SNe
Ia, and adopting the same luminosity width correction method (
$\Delta m_{15}$) for all of them, it shows the resulting
confidence regions corresponding to 68.3\%, 95.4\%, and 99.7\%
probability in the $(\Omega_M,\Omega_\Lambda)$-plane. Taken at
face value, this result excludes $\Omega_\Lambda = 0$ for values
of $\Omega_M$ which are consistent with other observations (e.g.,
of clusters of galaxies). This is certainly the case if a flat
universe is assumed. The probability regions are inclined along
$\Omega_\Lambda \approx 1.3 \Omega_M + (0.3\pm 0.2)$. It will turn
out  that this information is largely complementary to the
restrictions we shall obtain in Sect.6 from the CMB anisotropies.
 
\begin{figure}
\begin{center}
\includegraphics[height=0.6\textheight]{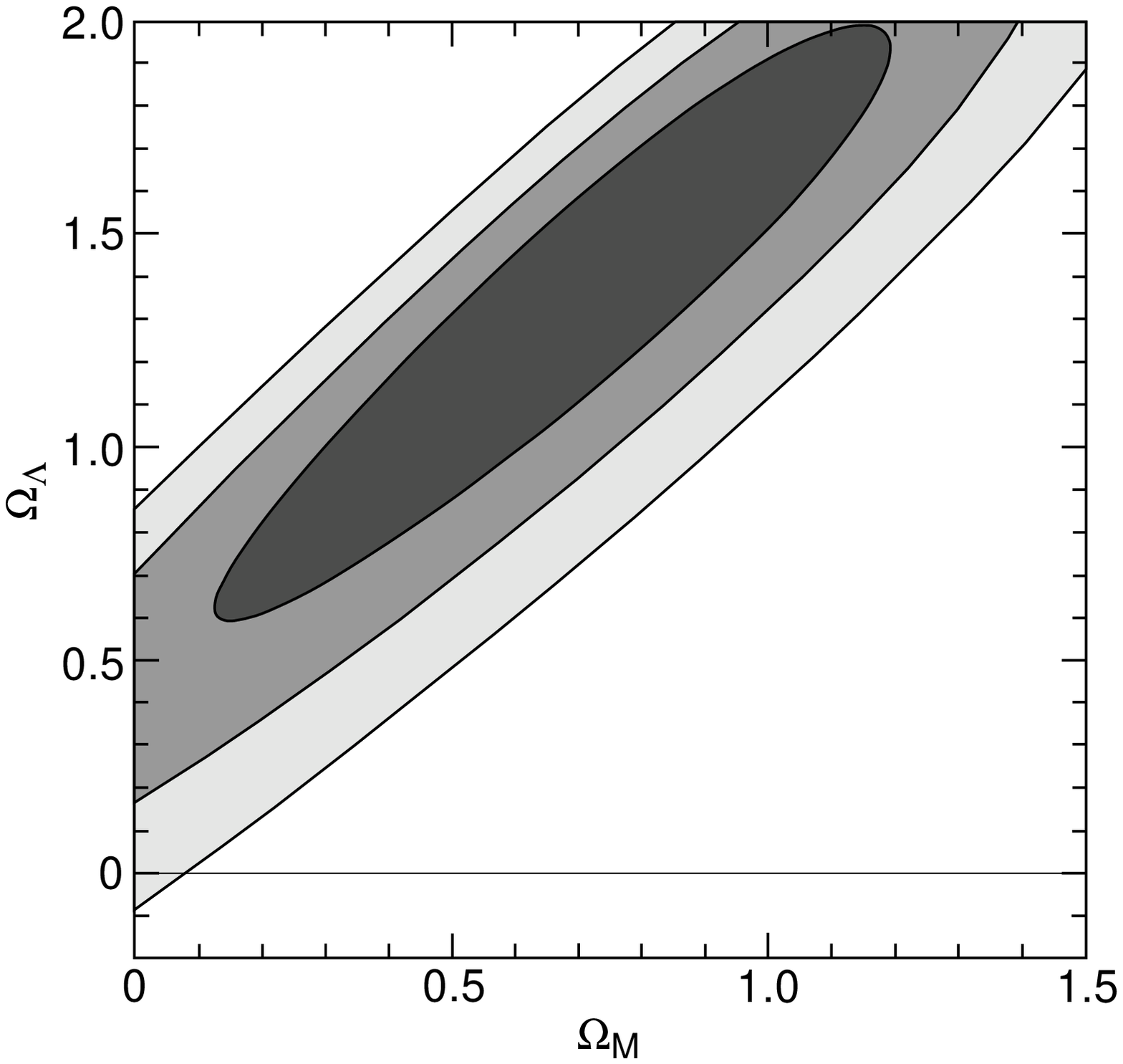}
\caption{Likelihood regions in the $\Omega_M - \Omega_\Lambda$ 
plane for the data in Fig.1. Contours give the 68.3\%, 95.4\%, and 99.7\% 
statistical confidence regions (adapted from \cite{47}).}
\label{Fig-2}
\end{center}
\end{figure}

\subsection{Systematic uncertainties}

Possible systematic uncertainties due to astrophysical effects
have been discussed extensively in the literature. The most
serious ones are (i) {\it dimming} by intergalactic dust, and (ii)
{\it evolution} of SNe Ia over cosmic time, due to changes in
progenitor mass, metallicity, and C/O ratio. I discuss these
concerns only briefly (see also \cite{47}, \cite{48}).

Concerning extinction, detailed studies show that high-redshift SN
Ia suffer little reddening; their B-V colors at maximum brightness
are normal. However, it can a priori not be excluded that we see
distant SNe through a grey dust with grain sizes large enough as
to not imprint the reddening signature of typical interstellar
extinction. One argument against this hypothesis is that this
would also imply a larger dispersion than is observed. The
discovery \cite{49} of SN 1997ff with the very high redshift
$z\approx 1.7$ led to the conclusion that its redshift and
distance estimates are inconsistent with grey dust. Perhaps this
statement is too strong, because a pair of galaxies in the
foreground of SN 1997ff at $z = 0.56$ may induce a magnification
due to gravitational lensing of $\sim 0.3 mag$ \cite{50}. With
more examples of this type the issue could be settled. Eq.(63)
shows that at redshifts $z\geq (2\Omega_\Lambda /\Omega_M)^{1/3}-1
\simeq 1.2$ the Universe is {\it decelerating}, and this provides an 
almost unambiguous signature for $\Lambda$, or some effective equivalent.

The same SN has provided also some evidence against a simple
luminosity evolution that could mimic an accelerating Universe.
Other empirical constraints are obtained by comparing subsamples
of low-redshift SN Ia believed to arise from old and young
progenitors. It turns out that there is no difference within the
measuring errors, {\it after} the correction based on the
light-curve shape has been applied. Moreover, spectra of
high-redshift  SNe appear remarkably similar to those at low
redshift. This is very reassuring. On the other hand, there seems 
to be a trend that more distant supernovas are bluer. It would, of 
course, be helpful if evolution could be predicted theoretically, but 
in view of what has been said earlier, this is not (yet) possible.

In conclusion, none of the investigated systematic errors appear
to reconcile the data with $\Omega_\Lambda = 0$ and $q_0\geq 0$.
But further work is necessary before we can declare this as a
really established fact.

To improve the observational situation a satellite mission called
SNAP (``Supernovas Acceleration Probe'') has been proposed
\cite{51}. According to the plans this satellite would observe
about 2000 SNe within a year and much more detailed studies could
then be performed. For the time being some scepticism with regard
to the results that have been obtained is not out of place.

Finally, I mention a more theoretical complication. In the
analysis of the data the luminosity distance for an ideal
Friedmann universe was always used. But the data were taken in the
real inhomogeneous Universe. This may not be good enough,
especially for high-redshift standard candles. The simplest way to
take this into account is to introduce a filling parameter which,
roughly speaking, represents matter that exists in galaxies but
not in the intergalactic medium. For a constant filling parameter
one can determine the luminosity distance by solving the
Dyer-Roeder equation. But now one has an additional parameter in
fitting the data. For a flat universe this was recently
investigated in \cite{52}.

\section{Microwave background anisotropies}

By observing the cosmic microwave background (CMB) we can directly
infer how the Universe looked at the time of recombination. Beside
its spectrum, which is Planckian to an incredible degree
\cite{53}, we also can study the temperature fluctuations over the
``cosmic photosphere'' at a redshift $z\approx1100$. Through these
we get access to crucial cosmological information (primordial
density spectrum, cosmological parameters, etc). A major reason
for why this is possible relies on the fortunate circumstance that
the fluctuations are tiny ($\sim 10^{-5}$ ) at the time of
recombination. This allows us to treat the deviations from
homogeneity and isotropy for an extended period of time
perturbatively, i.e., by linearizing the Einstein and matter
equations about solutions of the idealized Friedmann-Lema\^{\i}tre
models. Since the physics is effectively {\it linear}, we can
accurately work out the {\it evolution} of the perturbations
during the early phases of the Universe, given a set of
cosmological parameters. Confronting this with observations, tells
us a lot about the initial conditions, and thus about the physics
of the very early Universe. Through this window to the earliest
phases of cosmic evolution we can, for instance, test general
ideas and specific models of inflation.

\subsection{On the physics of CMB}

Long before recombination (at temperatures $T>6000 K$, say)
photons, electrons and baryons were so strongly coupled that these
components may be treated together as a single fluid. In addition
to this there is also a dark matter component. For all practical
purposes the two interact only gravitationally. The investigation
of such a two-component fluid for small deviations from an
idealized Friedmann behavior is a well-studied application of
cosmological perturbation theory. (For the basic equations and a
detailed analytical study, see \cite{54} and \cite{55}.)

At a later stage, when decoupling is approached, this approximate
treatment breaks down because the mean free path of the photons
becomes longer (and finally `infinite' after recombination). While
the electrons and baryons can still be treated as a single fluid,
the photons and their coupling to the electrons have to be
described by the general relativistic Boltzmann equation. The
latter is, of course, again linearized about the idealized
Friedmann solution. Together with the linearized fluid equations
(for baryons and cold dark matter, say), and the linearized
Einstein equations one arrives at a complete system of equations
for the various perturbation amplitudes of the metric and matter
variables. There exist widely used codes \cite{56}, \cite{57} that
provide the CMB anisotropies -- for given initial conditions -- to
a precision of about 1\%.

A lot of qualitative and semi-quantitative insight into the
relevant physics can be gained by looking at various
approximations of the `exact' dynamical system. Below I shall
discuss some of the main points. (For well-written papers on this
aspect I recommend \cite{58}, \cite{59}.)

For readers who want to skip this somewhat technical discussion 
and proceed directly to the observational results (Sect.6.2), the 
following qualitative remarks may be useful. A characteristic scale, 
which is reflected in the observed CMB anisotropies, is the sound 
horizon at last scattering, i.e., the distance over which a pressure 
wave can propagate until $\eta_{dec}$. This can be computed within the 
unperturbed model and subtends about one degree on the sky for typical
cosmological parameters. For scales larger than this sound horizon the
fluctuations have been laid down in the very early Universe. These have
been detected by the COBE satellite. The (brightness) temperature 
perturbation $\Theta = \Delta T/T$ (defined precisely in Eq.(88) below)
is dominated by the combination of the intrinsic temperature fluctuations
and gravitational redshift or blueshift effects. For example, photons that 
have to climb out of potential wells for high-density regions are redshifted. 
In Sect.6.1.5 it is shown that these effects combine for adiabatic initial
conditions to $\frac{1}{3}\Psi$, where $\Psi$ is the gravitational Bardeen 
potential (see Eq.(73)). The latter, in turn, is directly related to the density 
perturbations. For scale-free initial perturbations the corresponding angular
power spectrum of the temperature fluctuations turns out to be nearly flat
(Sachs-Wolfe plateau in Fig.3). The $C_l$ plotted in Fig.3 are defined in (109)
as the expansion coefficients of the angular correlation function in terms of
Legendre polynomials.

On the other hand, inside the sound horizon (for $\eta \leq \eta_{dec}$), acoustic,
Doppler, gravitational redshift, and photon diffusion effects combine to the 
spectrum of small angle anisotropies shown in Fig.3. These result from 
gravitationally driven acoustic oscillations of the photon-baryon fluid, which 
are damped by photon diffusion (Sect.6.1.4).

\subsubsection{Cosmological perturbation theory}

Unavoidably, the detailed implementation of what has just been
outlined is somewhat complicated, because we are dealing with
quite a large number of dynamical variables. This is not the place
to develop cosmological perturbation theory in any detail
\footnote{There is by now an extended literature on cosmological
perturbation theory. Beside the recent book \cite{60}, the review
articles \cite{61}, \cite{62}, and \cite{63} are recommended.
Especially \cite{61} is still useful for the general (gauge
invariant) formalism for multi-component systems. Unpublished
lecture notes by the author \cite{64} are planned to become
available.}, but I have to introduce some of it.

\subsection*{Mode decomposition}

Because we are dealing with slightly perturbed Friedmann
spacetimes we may regard the various perturbation amplitudes as
time dependent functions on a three-dimensional Riemannian space
$(\Sigma,\gamma)$ of constant curvature $K$. Since such a space is
highly symmetrical we are invited to perform two types of
decompositions.

In a first step we split the perturbations into {\it scalar,
vector}, and {\it tensor} contributions. This is based on the
following decompositions of vector and symmetric tensor fields on
$\Sigma$ : A vector field $\xi$ is a unique sum of a gradient and
a vector field $\xi_*$ with vanishing divergence,
\begin{equation}
\xi = \xi_* + \nabla f, \;\;\; \nabla \cdot \xi_* = 0.
\end{equation}
(If $\Sigma$ is noncompact we have to impose some fall-off
conditions.) The first piece $\xi_*$ is the `vector' part, and
$\nabla f$ is the `scalar' part of $\xi$. This is a special case
of the Hodge decomposition for differential forms. For a symmetric
tensor field $S_{ij}$ we have correspondingly :
\begin{equation}
S_{ij} = S_{ij}^{(scalar)} + S_{ij}^{(vector)} +S_{ij}^{(tensor)},
\end{equation}
with
\begin{eqnarray}
S_{ij}^{(scalar)}& = &\gamma_{ij}S^k{} _k + (\nabla_i \nabla_j -
\frac{1}{3}\gamma_{ij }\Delta )f, \\ S_{ij}^{(vector)}& =
&\nabla_i \xi_j + \nabla_j \xi_i,
\end{eqnarray}
with $\nabla_k \xi^k = 0$, and where $S_{ij}^{(tensor)}$ is a
symmetric tensor field with vanishing trace and zero divergence.

The main point is that these decompositions respect the covariant
derivative $\nabla$ on $(\Sigma,\gamma)$. For example, if we apply
the Laplacian on (64) we readily obtain
\[ \Delta\xi = \Delta\xi_* + \nabla(\Delta f + 2Kf),\]
and here the first term has vanishing divergence. For this reason
the different components in the perturbation equations {\it do not
mix}.

In a second step we can perform a {\it harmonic decomposition}, in
expanding all amplitudes in terms of generalized spherical
harmonics on $(\Sigma,\gamma)$. For $K=0$ this is just Fourier
analysis. Again the various modes do not mix, and very
importantly, the perturbation equations become for each mode {\it
ordinary} differential equations. (From the Boltzmann equation we
get an infinite hierarchy; see below.)

\subsection*{Gauge transformations, gauge invariant amplitudes}

In general relativity the diffeomorphism group of spacetime is an
{\it invariance group}. This means that the physics is not changed
if we replace the metric $g$ and all the matter variables
simultaneously by their diffeomorphically transformed objects. For
small amplitude departures from some unperturbed situation,
$g=g^{(0)}+\delta g$, etc., this implies that we have the {\it
gauge freedom}
\begin{equation}
\delta g \longrightarrow \delta g + L_\xi g^{(0)}, \; \;\;
etc.,\end{equation} where $L_\xi$ is the Lie derivative with
respect to any vector field $\xi$. Sets of metric and matter
perturbations which differ by Lie derivatives of their unperturbed
values are physically equivalent. Such gauge transformations
induce changes in the various perturbation amplitudes. It is
clearly desirable to write all independent perturbation equations
in a {\it manifestly gauge invariant} manner. Then one gets rid of
uninteresting gauge modes, and misinterpretations of the formalism
are avoided.

Let me show how this works for the metric. The most general {\it
scalar} perturbation $\delta g$ of the Friedmann metric
\begin{equation}
g^{(0)} = dt^2 - a^2(t)\gamma = a^2(t)\Bigl[d\eta^2 - \gamma
\Bigr]\end{equation} can be parameterized as follows
\begin{equation}
\delta g = 2a^2(\eta)\Bigl[ Ad\eta^2 + B_{,i}dx^id\eta -
(D\gamma_{ij} + E_{|ij} )dx^idx^j \Bigr].
\end{equation}
The functions $A(\eta,x^i),B,D,E$ are the scalar perturbation
amplitudes; $E_{|ij}$ denotes the second covariant derivative
$\nabla_i\nabla_j E$ on $(\Sigma,\gamma)$. It is easy to work out
how $A,B,D,E$ change under a gauge transformation (68) for a
vector field $\xi$ of `scalar' type: $\xi=\xi^0 \partial_0 + \xi^i
\partial_i,$  with $\xi^i=\gamma^{ij}\xi_{|j}$. From the result
one can see that the following {\it Bardeen potentials} \cite{61}
\begin{eqnarray}
\Psi &=& A - \frac{1}{a}\Bigl[ a(B + E')\Bigr]', \\
\Phi &=& D - \mathcal{H} (B + E')
\end{eqnarray}
are {\it gauge invariant}. Here, a prime denotes the derivative
with respect to the conformal time $\eta$, and $\mathcal{H}=a'/a$.
The potentials $\Psi$ and $\Phi$ are the only independent gauge
invariant metric perturbations of scalar type. One can always
chose the gauge such that only the $A$ and $D$ terms in (70) are
present. In this so-called {\it longitudinal} or {\it conformal
Newtonian} gauge we have $\Psi=A, \Phi=D$, hence the metric
becomes
\begin{equation}
g = a^2(\eta)\Bigl[ (1+2\Psi)d\eta^2 -
(1+2\Phi)\gamma_{ij}dx^idx^j \Bigr].
\end{equation}

\subsection*{Boltzmann hierarchy}

Boltzmann's description of kinetic theory in terms of a one
particle distribution function finds a natural setting in general
relativity. The metric induces a diffeomorphism between the
tangent bundle $TM$ and the cotangent bundle $T^*M$ over the
spacetime manifold $M$. With this the standard symplectic form on
$T^*M$ can be pulled back to $TM$. In natural bundle coordinates
the diffeomorphism is: $(x^\mu,p^\alpha)\mapsto (x^\mu,p_\alpha =
g_{\alpha\beta}p^\beta)$ , hence the symplectic form on $TM$ is
given by
\begin{equation}
\omega_g = dx^\mu\wedge d(g_{\mu\nu}p^\nu).
\end{equation}
The {\it geodesic spray} is the Hamiltonian vector field $X_g$ on
$(TM,\omega_g)$ belonging to the ``Hamiltonian function''
$L=\frac{1}{2}g_{\mu\nu}p^\mu p^\nu$. Thus, in standard notation,
\begin{equation}
i(X_g)\; \omega_g = dL.
\end{equation}
In bundle coordinates
\begin{equation}
X_g = p^\mu\frac{\partial}{\partial x^\mu} - \Gamma^\mu
{}_{\alpha\beta} p^\alpha p^\beta\frac{\partial}{\partial p^\mu}.
\end{equation}
The integral curves of this vector field satisfy the canonical
equations
\begin{eqnarray}
\frac{dx^\mu}{d\lambda} &= &p^\mu, \\  \frac{dp^\mu}{d\lambda} &
=& -\Gamma^\mu{}_{\alpha\beta} p^\alpha p^\beta.
\end{eqnarray}

The {\it geodesic flow} is the flow of $X_g$. The Liouville volume
form $\Omega_g$ is proportional to the fourfold wedge product of
$\omega_g$, and has the bundle coordinate expression
\begin{equation}
\Omega_g = (-g) dx^{0123} \wedge dp^{0123},
\end{equation}
where $dx^{0123} \equiv dx^0 \wedge dx^1 \wedge dx^2 \wedge dx^3$, etc... 
The one-particle phase space for particles of mass $m$ is the
submanifold $\Phi_m = \{ v\in TM \mid g(v,v) = m^2\}$ of $TM$.
This is invariant under the geodesic flow. The restriction of
$X_g$ to $\Phi_m$ will also be denoted by $X_g$. $\Omega_g$
induces a volume form $\Omega_m$ on $\Phi_m$, which is remains
invariant under $X_g$, thus $L_{X_g} \Omega_m = 0$. A simple
calculation shows that $\Omega_m = \eta\wedge\Pi_m$, where $\eta$
is the standard volume form of $(M,g)$, 
$\eta =\sqrt{-g}dx^{0123}$, and $\Pi_m = \sqrt{-g}dp^{123}/p_0,  p_0$ 
being determined by $g_{\mu\nu} p^\mu p^\nu = m^2$.
\\
Let $f$ be a distribution function on $\Phi_m$. The particle
number current density is
\begin{equation}
n^\mu(x) = \int_{P_m(x)} f p^\mu \Pi_m,
\end{equation}
where $P_m(x)$ is the fiber over $x$ in $\Phi_m$ (all momenta with
$g(p,p)=m^2)$. Similarly, the energy-momentum tensor is
\begin{equation}
T^{\mu\nu} = \int f p^\mu p^\nu \Pi_m.
\end{equation}
One can show that
\begin{equation}
\nabla_\mu n^\mu = \int_{P_m} (L_{X_g} f)\Pi_m,\end{equation} and
\begin{equation}
\nabla_\nu T^{\mu\nu} = \int_{P_m} p^\mu(L_{X_g}f)\Pi_m.
\end{equation}

The {\it Boltzmann equation} has the form
\begin{equation}
L_{X_g} f = C[f],
\end{equation}
where $C[f]$ is the collision term. If this is (for instance)
inserted into (83), we get an expression for the divergence of
$T^{\mu\nu}$ in terms of a collision integral. For collisionless
particles (neutrinos) this vanishes, of course.

Turning to perturbation theory, we set again $f = f^{(0)} + \delta
f$, where $f^{(0)}$ is the unperturbed distribution function of
the Friedmann model. For the perturbation $\delta f$ we choose as
independent variables $\eta, x^i, q, \gamma^j$, where $q$ is the
magnitude and the $\gamma^j$ the directional cosines of the
momentum vector relative to an orthonormal triad field $\hat{e}_i
(i=1,2,3)$ of the unperturbed spatial metric $\gamma$ on $\Sigma$.

>From now on we consider always the massless case (photons). By
investigating the gauge transformation behavior of $\delta f$
\cite{65} one can define a gauge invariant perturbation
$\mathcal{F}$ which reduces in the longitudinal gauge to $\delta
f$ (there are other choices possible \cite{65}), and derive with
some effort the following linearized Boltzmann equation for
photons:
\begin{eqnarray}
\lefteqn{ (\partial_\eta + \gamma^i \hat{e}_i)\mathcal{F} -
\hat{\Gamma}^i{}_{jk} \gamma^j \gamma^k
\frac{\partial\mathcal{F}}{\partial \gamma^i} - q\Bigl[ \Phi' +
\gamma^i \hat{e}_i \Psi \Bigr]\frac{\partial f^{(0)}}{\partial q}
 =} \nonumber \\ & & ax_e n_e \sigma_T \Bigl[ <\mathcal{F}> - \mathcal{F} -
q\frac{\partial f^{(0)}}{\partial q} \gamma^i \hat{e}_i V_b +
\frac{3}{4}Q_{ij} \gamma^i \gamma^j \Bigr].
\end{eqnarray}
On the left, the $ \hat{\Gamma}^i{}_{jk}$ denote the Christoffel
symbols of $(\Sigma,\gamma)$ relative to the triad  $\hat{e}_i$.
On the right, $x_e n_e$ is the unperturbed free electron density
($x_e = $ ionization fraction), $\sigma_T$ the Thomson cross
section, and $V_b$ the gauge invariant scalar velocity
perturbation of the baryons. Furthermore, we have introduced the
spherical averages
\begin{eqnarray}
<\mathcal{F}> &= & \frac{1}{4\pi}\int_{S^2} \mathcal{F}
d\Omega_\gamma, \\
Q_{ij} &=& \frac{1}{4\pi} \int_{S^2} [ \gamma_i \gamma_j -
\frac{1}{3}\delta_{ij}] \mathcal{F} d\Omega_\gamma.
\end{eqnarray}

In our applications to the CMB we work with the gauge invariant
{\it brightness temperature} perturbation
\begin{equation}
\Theta(\eta,x^i,\gamma^j) = \int \mathcal{F} q^3dq \; \Big / \;\;
4\int f^{(0)}q^3dq.
\end{equation}
(The factor $4$ is chosen because of the Stephan-Boltzmann law,
according to which $\delta \rho / \rho = 4 \delta T / T.$) It is
simple to translate (85) to the following equation for $\Theta$
\begin{eqnarray}
\lefteqn{(\Theta + \Psi)' + \gamma^i \hat{e}_i(\Theta + \Psi ) -
\hat{\Gamma}^i{}_{jk} \gamma^j \gamma^k \frac{\partial}{\partial
\gamma^i}(\Theta + \Psi ) = } \nonumber \\ & & (\Psi'-\Phi') + \dot{\tau}(\theta_0
-\Theta + \gamma^i \hat{e}_i V_b + \frac{1}{16}\gamma^i\gamma^j
\Pi_{ij}),\end{eqnarray} with $\dot{\tau}=x_e n_e \sigma_T a/a_0,\;
\theta_0=<\Theta>$ (spherical average),
\begin{equation}
\frac{1}{12} \Pi_{ij} = \frac{1}{4\pi}\int [\gamma_i\gamma_j -
\frac{1}{3}\delta_{ij} ] \Theta \; d\Omega_\gamma.
\end{equation}

Let me from now on specialize to the spatially flat case $(K=0)$.
In a mode decomposition (Fourier analysis of the
$x^{i}$-dependence), and introducing the {\it brightness moments}
$\theta_l(\eta)$ by
\begin{equation}
\Theta(\eta,k^i,\gamma^j) = \sum_{l=0}^{\infty }(-i)^l
\theta_l(\eta,k) P_l(\mu), \; \; \; \mu =
\mbox{\boldmath$\hat{k}\cdot\gamma$},
\end{equation}
we obtain
\begin{equation}
\Theta' + ik\mu(\Theta + \Psi) = -\Phi' +\dot{\tau}[ \theta_0 -
\Theta - i\mu V_b -\frac{1}{10}\theta_2 P_2(\mu)].
\end{equation}
It is now straightforward to derive from the last two equations
the following hierarchy of ordinary differential equations for the
brightness moments\footnote{In the literature the normalization of
the $\theta_l$ is sometimes chosen differently:
$\theta_l\rightarrow (2l+1)\theta_l$.} $\theta_l(\eta)$:
\begin{eqnarray}
\theta_0' & = &-\frac{1}{3}k \theta_1 - \Phi',  \\
\theta_1' & = & k\Bigl(\theta_0 + \Psi -\frac{2}{5}\theta_2 \Bigr)
-\dot{\tau}(\theta_1 -V_b),  \\
\theta_2' & = & k\Bigl(\frac{2}{3}\theta_1
-\frac{3}{7}\theta_3\Bigr) -
\dot{\tau} \frac{9}{10}\theta_2, \\
\theta_l' & = & k\Bigl(\frac{l}{2l-1}\theta_{l-1}
-\frac{l+1}{2l+3}\theta_{l+1}\Bigr),  \; \; \; l>2.
\end{eqnarray}

\subsection*{The complete system of perturbation equations}

Without further ado I collect below the complete system of
perturbation equations. For this some additional notation has to
be fixed.

Unperturbed {\it background} quantities: $\rho_\alpha, p_\alpha$
denote the densities and pressures for the species $\alpha = b$
(baryon and electrons), $\gamma$ (photons), $c$ (cold dark
matter); the total density is the sum $\rho = \sum_\alpha
\rho_\alpha$, and the same holds for the total pressure $p$. We
also use $w_\alpha = p_\alpha/\rho_\alpha, w=p/\rho $. The sound
speed of the baryon-electron fluid is denoted by $c_b$, and $R$ is
the ratio $3\rho_b/4\rho_\gamma $.

Here is the list of gauge invariant {\it scalar perturbation}
amplitudes (for further explanations see \cite{62}):
\begin{itemize}
\item
$\delta_\alpha, \delta$ : density perturbations ($\delta
\rho_\alpha /\rho_\alpha,  \delta \rho /\rho \;$  in the
longitudinal gauge); clearly: $\rho \; \delta = \sum \rho_\alpha
\delta_\alpha$.
\item
$V_\alpha ,V$ : velocity perturbations; $\rho(1+w)V=\sum_\alpha
\rho_\alpha (1+w_\alpha)V_\alpha.$
\item
$\theta_l, N_l$ : brightness moments for photons and neutrinos.
\item
$\Pi_\alpha, \Pi$ : anisotropic pressures;  $\Pi=\Pi_\gamma +
\Pi_\nu$. For the lowest moments the following relations hold:
\begin{equation}
\delta_\gamma = 4\theta_0, \; \; V_\gamma = \theta_1, \; \;
\Pi_\gamma = \frac{12}{5} \theta_2,
\end{equation}
and similarly for the neutrinos.
\item
$\Psi, \Phi $: Bardeen potentials for the metric perturbation.
\end{itemize}

As independent amplitudes we can choose: $\delta_b, \delta_c, V_b,
V_c, \Phi, \Psi, \theta_l, N_l$. The basic evolution equations
consist of three groups.
\begin{itemize}
\item
Fluid equations:
\begin{eqnarray}
\delta_c' & = & -kV_c - 3\Phi',  \\
V_c' &=& -aHV_c + k\Psi;  \\
\delta_b' &=& -kV_b - 3\Phi',  \\
V_b' &=& -aHV_b + kc^2_b \delta_b + k\Psi + \dot{\tau}(\theta_1
-V_b) / R .
\end{eqnarray}
\item
Boltzmann hierarchies for photons (Eqs. (93)-(96)) and the
collisionless neutrinos.
\item
Einstein equations : We only need the following algebraic ones for
each mode:
\begin{eqnarray}
k^2\Phi &=& 4\pi G a^2 \rho\Bigl[ \delta +
3\frac{aH}{k}(1+w)V\Bigr ], \\
k^2(\Phi + \Psi) &=& -8\pi G a^2 p \; \Pi.
\end{eqnarray}
\end{itemize}

In arriving at these equations some approximations have been made
which are harmless \footnote{In the notation of \cite{62} we have
set $q_\alpha = \Gamma_\alpha =0$, and are thus ignoring certain
intrinsic entropy perturbations within individual components.},
except for one: We have ignored polarization effects in Thomson
scattering. For quantitative calculations these have to be
included. Moreover, polarization effects are highly interesting,
as I shall explain later.

\subsubsection{Angular correlations of temperature fluctuations}

The system of evolution equations has to be supplemented by
initial conditions. We can not hope to be able to predict these,
but at best their statistical properties (as, for instance, in
inflationary models). Theoretically, we should thus regard the
brightness temperature perturbation $\Theta(\eta,x^i,\gamma^j)$ as
a random field. Of special interest is its angular correlation
function at the present time $\eta_0$. Observers measure only one
realization of this, which brings unavoidable {\it cosmic
variances}.

For further elaboration we insert (91) into the Fourier expansion
of $\Theta$, obtaining
\begin{equation}
\Theta(\eta,\mbox{\boldmath$x$},\mbox{\boldmath$\gamma$}) =
(2\pi)^{-\frac{3}{2}} \int d^3k\sum_l \theta_l(\eta,k)
G_l(\mbox{\boldmath$x$},\mbox{\boldmath$\gamma$};\mbox{\boldmath$k$}),
\end{equation}
where
\begin{equation}
G_l(\mbox{\boldmath$x$},\mbox{\boldmath$\gamma$};\mbox{\boldmath$k$})
= (-i)^l P_l(\mbox{\boldmath$\hat{k}\cdot \gamma$})
\exp(i\mbox{\boldmath$k\cdot x$}).
\end{equation}
Hence we have
\begin{equation}
\Theta(\eta,\mbox{\boldmath$x$},\mbox{\boldmath$\gamma$}) =
\sum_{lm}a_{lm}^* Y_{lm}(\mbox{\boldmath$\gamma$}),
\end{equation}
with
\begin{equation}
a_{lm} = (2\pi)^{-\frac{3}{2}}\int d^3k \;\theta_l(\eta,k)\; i^l
\frac{4\pi}{2l+1}Y_{lm}(\mbox{\boldmath$\hat{k}$})\exp(-i\mbox{\boldmath$k\cdot
x$}).
\end{equation}
We expect on the basis of rotation invariance that the two-point
correlation of the random variables $a_{lm}$ has the form
\begin{equation}
<a_{lm} a_{l'm'}^*> = C_l \delta_{ll'} \delta_{mm'}.
\end{equation}
>From (106) and (108) we see that the angular correlation function
of $\Theta$ in \textbf{x}-space is
\begin{equation}
<\Theta(\mbox{\boldmath$\gamma$})
\Theta(\mbox{\boldmath$\gamma'$})> = \sum_l \frac{2l+1}{4\pi}C_l
P_l(\mbox{\boldmath$\gamma\cdot\gamma'$}).
\end{equation}
If different modes in \textbf{k}-space are uncorrelated, we obtain
from (107)
\begin{equation}
\frac{2l+1}{4\pi}C_l = \frac{1}{2\pi^2}\int_0^\infty
\frac{dk}{k}\frac{k^3|\theta_l(k)|^2}{2l+1}.
\end{equation}

\subsubsection*{Cosmic variance}

The $C_l$ are the expectation values of the stochastic variable
\[ Z=\frac{1}{2l+1}\sum_m a_{lm}a_{lm}^* .\]
If the $a_{lm}$ are Gaussian random variables, as in simple inflationary
models, then the variance of $Z$, and thus of $C_l$, is easily
found to be given by
\begin{equation}
\sigma(C_l) = \sqrt{\frac{2}{2l+1}}.
\end{equation}
This is a serious limitation for low multipoles that cannot be
overcome. For large $l$ the measured $C_l$ should be accurately 
described by (110), taken at the present time.

\subsubsection{Brightness moments in sudden decoupling }

The linearized Boltzmann equation in the form (92) as an inhomogeneous 
linear differential for the $\eta$-dependence has the `solution'
\begin{eqnarray}
\lefteqn{ (\Theta + \Psi)(\eta_0,\mu;k)= }\nonumber \\ & &
\int_0^{\eta_0} d\eta\Bigl[ \dot{\tau}(\theta_0 + \Psi -i\mu V_b -
\frac{1}{10}\theta_2 P_2)+ \Psi'-\Phi'\Bigr ] 
 e^{-\tau(\eta,\eta_0)} e^{ik\mu(\eta-\eta_0)},
\end{eqnarray}
where
\begin{equation}
\tau(\eta,\eta_0) = \int_\eta^{\eta_0} \dot{\tau} d\eta
\end{equation}
is the {\it optical depth}. The combination $\dot{\tau}e^{-\tau}$ is the
(conformal) {\it time visibility function}. It has a simple 
interpretation: Let $p(\eta,\eta_0)$ be the probability
that a photon did not scatter between $\eta$ and today ($\eta_0$). Clearly, 
$p(\eta-d\eta,\eta_0) = p(\eta,\eta_0)(1-\dot{\tau}d\eta)$. Thus 
$p(\eta,\eta_0) = e^{-\tau(\eta,\eta_0)}$, and the visibility 
function times $d\eta$ is the probability that a photon last scattered between
$\eta$ and $\eta+d\eta$. The visibility function is therefore {\it strongly 
peaked} near decoupling. This is very useful, both for analytical and 
numerical purposes.

In order to obtain an integral representation for the multipole moments
$\theta_l$, we insert in (112) for the $\mu$-dependent factors standard 
expansions in terms of Legendre polynomials. For $l\geq 2$ we find the 
following useful formula:
\begin{equation} \frac{
\theta_l(\eta_0)}{2l+1} = \int_0^{\eta_0} d\eta e^{-\tau(\eta)}\Bigl[ (\dot{\tau}
\theta_0 + \dot{\tau}\Psi +\Psi'-\Phi' )j_l(k(\eta_0-\eta))  
 +\dot{\tau}V_b j_l' + \dot{\tau}\frac{1}{20}\theta_2(3j_l''+ j_l)\Bigr ].
\end{equation}

In a reasonably good approximation we can replace the visibility function
by a $\delta$-function, and obtain (with $\Delta\eta\equiv \eta_0 -\eta_{dec}, 
V_b(\eta_{dec})\simeq \theta_1(\eta_{dec})$:
\begin{equation}
\frac{\theta_l(\eta_0,k)}{2l+1}\simeq[ \theta_0 +\Psi](\eta_{dec},k)
j_l(k\Delta \eta) + \theta_1(\eta_{dec},k)j_l'(k\Delta\eta) + ISW + Quad.
\end{equation}
Here, the quadrupole contribution (last term) is not important. ISW denotes the 
{\it integrated Sachs-Wolfe effect}:
\begin{equation}
ISW = \int_{\eta_{dec}}^{\eta_0} d\eta(\Psi'-\Phi')j_l(k\Delta \eta),
\end{equation}
which only depends on the time variations of the Bardeen potentials 
between recombination and the present time.

The interpretation of the first two terms in (115) is quite obvious: 
The first describes the fluctuations of the {\it effective} 
temperature $\theta_0 +\Psi$ on the cosmic photosphere, as we would see
them for free streaming between there and us, -- if the gravitational 
potentials would not change in time. ($\Psi$ includes blue- and redshift
effects.) The dipole term has to be interpreted, of course, as a Doppler 
effect due to the velocity of the baryon-photon fluid.

In this approximate treatment we only have to know the effective 
temperature $\theta_0 + \Psi$ and the velocity moment $\theta_1$
at decoupling. The main point is that eq.(115) provides a good 
understanding of the physics of the CMB anisotropies. Note that the 
individual terms are all gauge invariant. In gauge dependent methods 
interpretations would be ambiguous.

\subsubsection{Acoustic oscillations}

In this subsection we derive from the Boltzmann hierarchy (93)-(96) 
an approximate equation for the effective temperature fluctuation 
$\Delta T\equiv\theta_0 + \Psi$, which will teach us a lot.

As long as the mean free path of photons is much shorter than the 
wavelength of the fluctuation, the optical depth through a wavelength
$\sim\dot{\tau}/k$ is large. Thus the evolution equations may be expanded 
in $k/\dot{\tau}$.

In lowest order we obtain $\theta_1 = V_b, \; \theta_l = 0$ for $l\geq 2$, thus 
$\delta_b' = 3\theta_0'$. Going to the first order, we can replace on 
the right of the following form of eq.(94),
\begin{equation}
\theta_1 - V_b = \frac{R}{\dot{\tau}}[V_b' + \frac{a'}{a}V_b - k\Psi],
\end{equation}
$V_b$ by $\theta_1$:
\begin{equation}
\theta_1 - V_b = \frac{R}{\dot{\tau}}[\theta_1' + \frac{a'}{a}\theta_1 -k\Psi].
\end{equation}
We insert this in (94), and set in first order  $\theta_2=0$. Using
also $a'/a = R'/R$ we get
\begin{equation}
\theta_1' = \frac{1}{1+R}k\theta_0 + k\Psi - \frac{R'}{1+R}\theta_1.
\end{equation}
Together with (93) we find the driven oscillator equation
\begin{equation}
\theta_0'' + \frac{R'}{1+R}\frac{a'}{a}\theta_0' + c_s^2 k^2 \theta_0 = F(\eta),
\end{equation}
where
\begin{equation}
c_s^2 = \frac{1}{3(1+R)}, \; \; F(\eta) = -\frac{k^2}{3}\Psi -
\frac{R}{1+R}\frac{a'}{a}\Phi' - \Phi''.
\end{equation}
The damping term is due to expansion. In second order one finds an 
additional damping term proportional to $\theta_0'$:
\begin{equation}
\frac{1}{3}\frac{k^2}{\dot{\tau}}\Bigl[(\frac{R}{1+R})^2 +\frac{8}{9}\frac{1}{1+R}
\Bigr]\theta_0'.
\end{equation}
This describes the damping due to photon diffusion (Silk damping).

We discuss here only the first order equation, which we rewrite in
the more suggestive form ($m_{eff} \equiv 1+R $)
\begin{equation}
(m_{eff}\theta_0')' + \frac{k^2}{3}(\theta_0 + m_{eff}\Psi) =  
-(m_{eff}\Phi')'.
\end{equation}

This equation may be interpreted as follows: The change in momentum 
of the photon-baryon fluid is determined by a competition between
pressure restoring and gravitational driving forces.

Let us, in a first step, ignore the time dependence of $m_{eff}$ (i.e., of the
baryon-photon ratio $R$), then we get a forced harmonic oscillator equation
\begin{equation}
m_{eff}\theta_0'' + \frac{k^2}{3}\theta_0 = -\frac{k^2}{3}m_{eff}\Psi -
(m_{eff}\Phi')'.
\end{equation}
The effective mass $m_{eff}=1+R$ accounts for the inertia of baryons.
Baryons also contribute gravitational mass to the system, as is evident 
from the right hand side of the last equation. Their contribution to the 
pressure restoring force is, however, negligible.

We now ignore in (124) also the time dependence of the gravitational
potentials $\Phi,\Psi$. With (121) this then reduces to
\begin{equation}
\theta_0'' + k^2 c_s^2 \theta_0 = -\frac{1}{3}k^2\Psi.
\end{equation}
This simple harmonic oscillator under constant acceleration provided 
by gravitational infall can immediately be solved:
\begin{equation}
\theta_0(\eta) = [ \theta_0(0) + (1+R)\Psi]\cos(kr_s) +
\frac{1}{kc_s}\dot{\theta}_0(0)\sin(kr_s) - (1+R)\Psi,
\end{equation}
where $r_s(\eta)$ is the comoving sound horizon $\int c_s d\eta$. 
One can show that for {\it adiabatic} initial conditions there is only 
a cosine term. In this case we obtain for $\Delta T$:
\begin{equation}
\Delta T(\eta,k) = [ \Delta T(0,k) + R\Psi]\cos(kr_s(\eta)) -R\Psi.
\end{equation}

\subsubsection*{Discussion}  
In the radiation dominated phase ($R=0$) this reduces to $\Delta T(\eta)
 \propto \cos kr_s(\eta)$, 
which shows that the oscillation of $\theta_0$ is 
displaced by gravity. The zero point corresponds to the state at which
gravity and pressure are balanced. The displacement $-\Psi>0$ yields hotter 
photons in the potential well since gravitational infall not only increases
the number density of the photons, but also their energy through gravitational
blue shift. However, well after last scattering the photons also suffer 
a redshift when climbing out of the potential well, which precisely
cancels the blue shift. Thus the effective temperature perturbation we
see in the CMB anisotropies is -- as remarked in connection with eq. (115)
-- indeed $\Delta T = \theta_0 + \Psi$.
It is clear from (127) that a characteristic wave-number is 
$k=\pi/r_s(\eta_{dec})\simeq \pi/c_s\eta_{dec}$. 
A spectrum of $k$-modes 
will produce a sequence of peaks with wave numbers 
\begin{equation}
k_m = m\pi/ r_s(\eta_{dec}), \; \; m = 1,2,...
\end{equation}
Odd peaks correspond to the compression phase (temperature crests), whereas 
even peaks correspond to the rarefaction phase (temperature troughs) inside
the potential wells. Note also that the characteristic length scale 
$r_s(\eta_{dec})$, which is reflected in the peak structure, is determined
by the underlying unperturbed Friedmann model. This comoving sound horizon
at decoupling depends on cosmological parameters, but not on 
$\Omega_\Lambda$. Its role will further be discussed in Sect.6.2. 
Inclusion of baryons not only changes the sound speed, but gravitational 
infal leads to greater compression of the fluid in a potential well, and thus 
to a further displacement of the oscillation zero point (last term in(127). 
This is not compensated by the redshift after last scattering, since the 
latter is not affected by the baryon content. As a result all peaks from
compression are enhanced over those from rarefaction. Hence, the relative 
heights of the first and second peak is a sensitive measure of the baryon 
content. We shall see that the inferred baryon abundance from the present
observations is in complete agreement with the results from big bang
nucleosynthesis.

What is the influence of the slow evolution of the effective mass 
$m_{eff}=1+R$? Well, from the adiabatic theorem we know that for a slowly
varying $m_{eff}$ the ratio energy/frequency is an adiabatic invariant.
If $A$ denotes the amplitude of the oscillation, the energy is 
$\frac{1}{2}m_{eff}\omega^2 A^2$. According to (121) the frequency 
$\omega=kc_s$ is proportional to $m_{eff}^{-1/2}$. Hence $A\propto\omega^{-1/2}
\propto m_{eff}^{1/4}\propto(1+R)^{-1/4}$.

\subsubsection{Angular power spectrum for large scales}
The {\it angular power spectrum} is defined as $l(l+1)C_l$ versus $l$. For large scales, 
i.e., small $l$, observed with COBE, the first term in eq.(115) dominates.
Let us have a closer look at this so-called Sachs-Wolfe contribution.

For large scales (small $k$) we can neglect in the first equation (93) of the 
Boltzmann hierarchy the term proportional to 
$k$: $\theta_0'\approx - \Phi'\approx \Psi'$, 
neglecting also $\Pi$ (i.e., $\theta_2$) on large scales. Thus 
\begin{equation}
\theta_0(\eta)\approx \theta_0(0) + \Psi(\eta) - \Psi(0).
\end{equation}
To proceed, we need a relation between $\theta_0(0)$ and $\Psi(0)$. This
can be obtained by looking at superhorizon scales in the tight coupling
limit. (Alternatively, one can investigate the Boltzmann hierarchy in
the radiation dominated era.) For adiabatic initial perturbations one easily finds 
$\theta_0(0)=-\frac{1}{2}\Psi(0)$, 
while for the isocurvature case one gets 
 $\theta_0(0)=\Psi(0)=0$. 
Using this in (129), and also that $\Psi(\eta)=\frac{9}{10}
\Psi(0)$ 
in the matter dominated era, we find for the effective temperature 
fluctuations at decoupling
\begin{equation}
(\theta_0 +\Psi)(\eta_{dec}) = \frac{1}{3}\Psi(\eta_{dec})
\end{equation}
for the adiabatic case. For initial isocurvature fluctuations the result is 
{\it six times larger}. Eq.(130) is due to Sachs and Wolfe. It allows 
us to express the angular CMB power spectrum on large scales in terms of the
power spectrum of density fluctuations at decoupling. If the latter has evolved 
from a scale free primordial spectrum, it turns out that $l(l+1)C_l$ is {\it
constant} for small $l's$. It should be emphasized that on these large scales
the power spectrum remains close to the primordial one.

Having discussed the main qualitative aspects, we show in Fig.3 a typical 
theoretical CMB power spectrum for scale free adiabatic initial conditions.

\begin{figure}
\begin{center}
\includegraphics{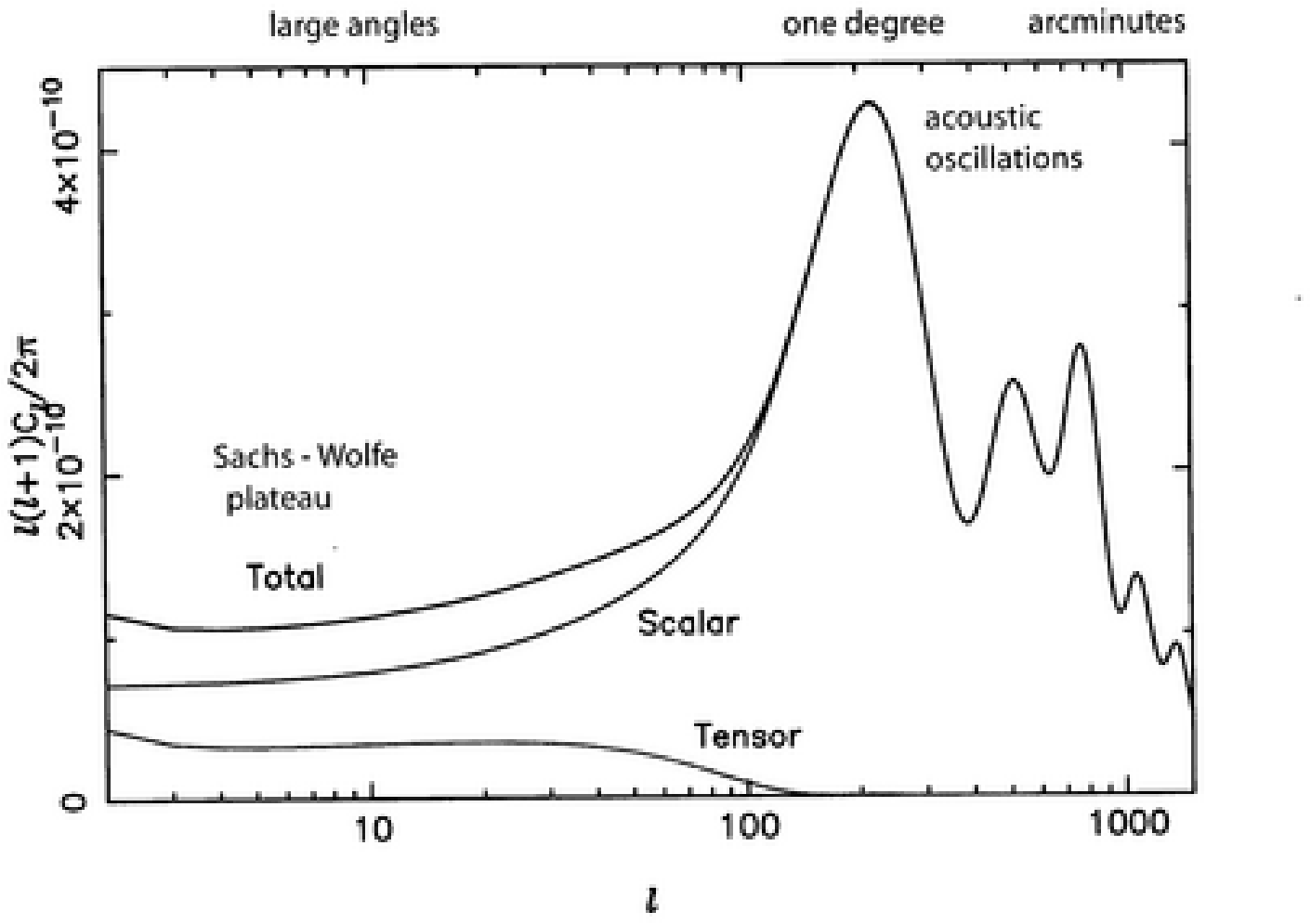}
\caption{Theoretical angular power spectrum for adiabatic initial 
perturbations and typical cosmological parameters. 
The scalar and tensor contributions to the anisotropies are also shown.} 
\label{Fig-3}
\end{center}
\end{figure}

\subsection{Observational results}

CMB anisotropies had been looked for ever since Penzias and Wilson's discovery
of the CMB, but had eluded all detection until the {\it Cosmic Background Explorer}
(COBE) satellite discovered them on large angular scales in 1992 \cite{66}. It is not at 
all astonishing that it took so long in view of the fact that the temperature
perturbations are only one part in $10^{-5}$ (after subtraction of the obvious 
dipole anisotropy). There are great experimental difficulties to isolate the
cosmologically interesting signal from foreground contamination. The most
important of these are: (i) galactic dust emission; (ii) galactic thermal
and synchrotron  emission; (iii) discrete sources; (iv) atmospheric emission,
in particular at frequencies higher than $\sim$10 GHz.

After 1992 a large number of ground and balloon-borne experiments
were set up to measure the anisotropies on smaller scales. Until quite 
recently the measuring errors were large and the data had a considerable
scatter, but since early 2001 the situation looks much better. Thanks to 
the experiments BOOMERanG \cite{67}, MAXIMA \cite{68} and DASI \cite{69}
we now have clear evidence
for multiple peaks in the angular power spectrum at positions and relative
heights that were expected on the basis of the simplest inflationary models 
and big bang nucleosynthesis.

Wang et al. 
\cite{70} have compressed all available data into a single band-averaged set of 
estimates of the CMB power spectrum. Their result, together with the 
$\pm 1\sigma $ errors, is reproduced in Fig.4. These data provide tight 
constraints for the cosmological 
parameters. However, the CMB anisotropies alone do not fix them all because there are 
unavoidable degeneracies, especially when tensor modes (gravity waves) are included. 
This degeneracy is illustrated in Fig.9 of Ref.[70] by three best fits that are obtained 
by fixing $\Omega_b h_0^2$ in a reasonable range.

\begin{figure}
\begin{center}
\includegraphics[height=0.7\textheight]{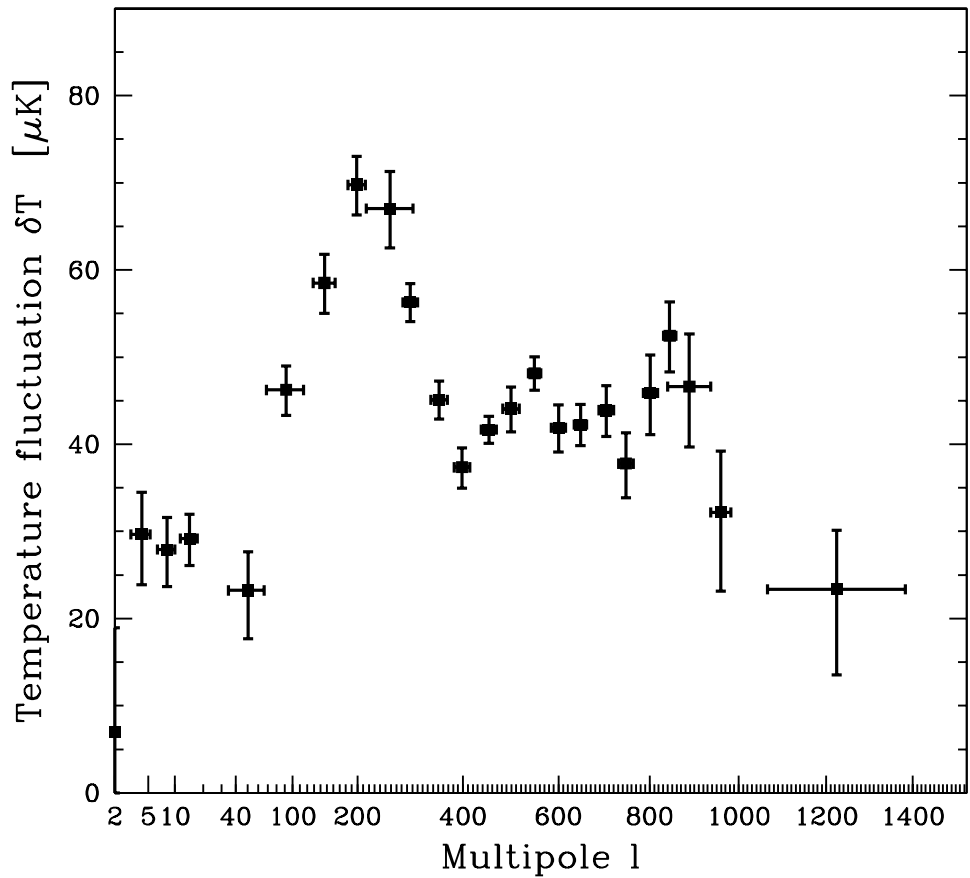}
\caption{ Band-averaged CMB power spectrum, together with the $\pm 1\sigma$
errors (Fig.2 of Ref. [70]).}
\label{Fig-4}
\end{center}
\end{figure}

Such degeneracies can only be lifted if other cosmological information is used.
Beside the supernova results, discussed in Sect.5, use has been made, for instance,
of the available information for the galaxy power spectrum. In \cite{71} the CMB data 
have been combined with the power spectrum of the $2dF$ (2-degree-Field) 
Galaxy Redshift Survey (2dFGRS). The authors summarize their results of the combined 
likelihood analysis in Table 1 of their paper. Here, I quote only part of it. The Table 
below shows the $\pm 2\sigma$ parameter ranges for some of the cosmological parameters, 
for two types of fits. In the first only the CMB data are used (but tensor modes are 
included), while in the second these data are combined with the 2dFGRS power spectrum 
(assuming adiabatic, Gaussian initial conditions described by power laws).

\vspace{10 mm}

\begin{tabular}{|l||c|c|} 
\multicolumn{3}{c}{\textbf{Table 1} } \\  \hline
Parameter & CMB alone & CMB and 2dFGRS\\ \hline\hline
$\Omega_b h_0^2$ & 0.016-0.045 & 0.018-0.034 \\
$\Omega_c h_0^2$ & 0.03-0.18   & 0.07-0.13   \\
$\Omega_K $      & -0.68-0.06  & -0.05-0.04  \\
$\Omega_\Lambda$  & $<$0.88       & 0.65-0.85   \\  \hline
\end{tabular}

\vspace{10 mm}
Note that $\Omega_K$ is not strongly constraint by CMB alone. However, if 
$h_0$ is assumed a priori to be within a reasonable range, then $\Omega_K$ 
has to be close to zero (flat universe). 
It is very satisfying that the combination of the CMB and 2dFGRS data constrain 
$\Omega_\Lambda$ in the range $0.65 \leq   \Omega_\Lambda \leq 0.85$. This 
is independent of -- but consistent with -- the supernova results.

Another beautiful result has to be stressed. For the baryon parameter
$\Omega_b h_0^2$ there is now full agreement between the CMB results 
and the BBN prediction. Earlier speculations in connection with possible
contradictions now have evaporated. The significance of this consistency cannot
be overemphasized.

All this looks very impressive. It is, however, not forbidden to still worry
about possible complications located in the initial conditions, for which 
we have no established theory. For example, an isocurvature admixture cannot 
be excluded and the primordial power spectrum may have unexpected features.

Temperature measurements will not allow us to isolate the contribution of 
gravitational waves. This can only be achieved with future sensitive polarization
experiments. Polarization information will provide crucial clues about the physics
of the very early Universe. It can, for instance, be used to discriminate between 
models of inflation. With the {\it Planck} satellite, currently scheduled for 
launch in February 2007, it will be possible to detect gravitational waves even 
if they contribute only 10 percent to the anisotropy signal.

\section{Quintessence}

For the time being, we have to live with the mystery of the incredible smallness of a
gravitationally effective vacuum energy density. For most physicists it is too much 
to believe that the vacuum energy constitutes the missing two thirds of the average
energy density of the {\it present} Universe. This would really be bizarre. The goal
of quintessence models is to avoid such an extreme fine-tuning. In many ways people
thereby repeat what has been done in inflationary cosmology. The main motivation there was, 
as is well-known, to avoid excessive fine tunings of standard big bang cosmology 
(horizon and flatness problems). 

In concrete models the exotic missing energy with negative pressure is again described
by a scalar field, whose potential is chosen such that the energy density  of the
homogeneous scalar field adjusts itself to be comparable to the matter density today
for quite generic initial conditions, and is dominated by the potential energy. This 
ensures that the pressure becomes sufficiently negative. It is not simple to implement
this general idea such that the model is phenomenologically viable. For instance, the 
success of BBN should not be spoiled. CMB and large scale structure impose other 
constraints. One also would like to understand why cosmological acceleration 
started at about $z\sim 1$, and not much earlier or in the far future. There have
been attempts to connect this with some characteristic events in the post-recombination 
Universe. On a fundamental level, the origin of a quintessence field that must be 
extremely weakly coupled to ordinary matter, remains in the dark.

There is already an extended literature on the subject. Refs. \cite{72} - \cite{78} give 
a small selection of important early papers and some recent reviews. I conclude by 
emphasizing again that on the basis of the vacuum energy problem we would expect a 
huge additive constant for the quintessence potential that would destroy the hole 
picture. Thus, assuming for instance that the potential approaches zero as the scalar
field goes to infinity, has (so far) no basis. Fortunately, future more precise
observations will allow us to decide whether the presently dominating exotic energy
density satisfies $p/\rho = -1$ or whether this ratio is somewhere between $-1$ and $-1/3$.
Recent studies (see \cite{79}, \cite{80}, and references therein) which make use of existing
CMB data, SN Ia observations and other information do not yet support quintessence
with $w_Q > -1$.

However, even if convincing evidence for this should be established, we will not be able
to predict the distant future of the Universe. Eventually, the quintessence energy
density may perhaps become {\ negative}. This illustrates that we may never be able
to predict the asymptotic behavior of the most grandiose of all dynamical systems.
Other conclusions are left to the reader.

\end{document}